\documentclass[aps,prb,amsmath,amssymb,twocolumn,amsfonts,longbibliography]{revtex4-2}
\usepackage{amssymb}
\usepackage{latexsym}
\usepackage[colorlinks,allcolors=red]{hyperref}
\usepackage{epsfig}

\usepackage{dcolumn}
\usepackage{amsmath}
\usepackage{amsfonts}
\usepackage{bm}
\usepackage{color}
\usepackage{ulem}
\usepackage{comment}

\newcommand{\be}{\begin{equation}}
\newcommand{\ee}{\end{equation}}
\newcommand{\bea}{\begin{eqnarray}}
\newcommand{\eea}{\end{eqnarray}}

\newcommand{\p}{\partial}
\newcommand{\s}{\sigma}

\newcommand{\la}{\langle}
\newcommand{\ra}{\rangle}
\newcommand{\rd}{\mbox{d}}
\newcommand{\ri}{\mbox{i}}
\newcommand{\re}{\mbox{e}}

\newcommand{\dg}{{^\dagger}}

\renewcommand{\vec}[1]{{\bm #1}}

\begin{document}
\title{Tractable model for a fractionalized Fermi liquid (FL$^*$) on
a square lattice}
\author{Piers Coleman}
\affiliation{
Center for Materials Theory, Department of Physics and Astronomy,
Rutgers University, 136 Frelinghuysen Rd., Piscataway, NJ 08854-8019, USA}
\affiliation{Department of Physics, Royal Holloway, University
of London, Egham, Surrey TW20 0EX, UK.}
\author{Elio J. K\"onig}
\affiliation{Department of Physics, University of Wisconsin-Madison, Madison, Wisconsin 53706, USA}
\author{Aaditya Panigrahi}
\affiliation{Department of Physics, Cornell University, Ithaca, NY 14853, USA}
\author{Alexei Tsvelik}
\affiliation{Division of Condensed Matter Physics and Materials
Science, Brookhaven National Laboratory, Upton, NY 11973-5000, USA}

\date{\today}
\begin{abstract}
Motivated by the continued interest in Fermi-surface reconstruction without symmetry breaking, we present an analytically tractable microscopic model of a fractionalized Fermi liquid (FL$^*$) on a square lattice and discuss its potential relevance to the cuprates. As in ancilla-qubit constructions, the model is related to Kondo lattice systems, but in this case, the conduction electrons interact with a $\mathbb{Z}_2$ spin liquid of the Yao--Lee type, with a Majorana Fermi surface. The associated $\mathbb Z_2$ gauge theory is static so that the model can be analytically solved to leading-logarithic accuracy. There are two phases: one in which the fractionalized fermions of the spin liquid hybridize with  conduction electrons to form a common Fermi surface violating the naive Luttinger count, and one in which they remain decoupled. We discuss the salient features of the small Fermi-surface phase, including analytically derived momentum dependent coherence factors responsible for the appearance of Fermi arcs \`{a} la Yang-Rice-Zhang. We further discuss the impact of quantum and thermal fluctuations, including a strong diamagnetic response and a logarithmically divergent Sommerfeld coefficient at the onset of the pseudogap.
\end{abstract}

\maketitle

\section{Introduction}

The concept of a fractionalized Fermi liquid (FL$^*$), in which an electron Fermi liquid 
and a spin liquid
co-exist, was first proposed as a way to satisfy the Oshikawa sum rule at the large-to-small Fermi surface transition of a Kondo lattice\cite{senthil03}. Recently, this idea has re-emerged in the context of high temperature cuprate superconductors\cite{sachdev25}, to account for the identification of small Fermi-surface pockets in the underdoped compounds. The phase diagram of the cuprate superconductors is the intellectual proving-ground of a forty year old debate about the pairing mechanism and the strange normal state of these quantum materials. If the underdoped phase of these systems can indeed be understood as an FL$^*$, then a vital step-forward in our understanding will  have been made. 

Early evidence for Fermi pockets  came from angle-resolved photoemission studies (ARPES) which suggested that the features identified as Fermi-arcs in underdoped cuprates should be re-interpreted as  Fermi pockets in which the back-side of the pocket has a small ARPES coherence factor, making it almost invisible and disguising it as an arc\cite{Peter1}. 
New support for pockets is provided by 
the analysis of the of the angle-resolved magnetoresistance\cite{harrison25,Yamaji2025}, which displays an angular maximum or saddle point called a "Yamaji effect" { and from the observations of the quantum oscillations \cite{deHaas1,deHaas2}}. In the re-analysis, these features correspond to Fermi pockets with an area which scales with the hole density, rather than the electron density of a  Fermi liquid. The persistence of the angle-dependent resistivity to high temperatures rules out a doubling of the unit cell due to density wave formation, thus suggest the existence of a phase with Fermi surfaces that do not enclose the conventional Luttinger volume { which we define as} an FL$^*$.

Several theoretical proposals\cite{senthil, senthil2, Ashvin, Grover, Bonderson, Tsvelik2016} have posited that an FL$^*$ phase involves the co-existence of a charged Fermi liquid and a background spin liquid.  
  Here, the key idea is that one component of the electronic fluid has localized into a spin liquid which is invisible to ARPES experiments, so the naive Luttinger count of the electronic Fermi surface deviates from the true electron density.

Two dimensional spin liquids pose a theoretical challenge to condensed matter physics.  The simplest analytic approach to their description adopts an Abrikosov fermion (parton) decomposition of the spin operators, carrying out a mean-field treatment of the intersite Heisenberg interactions to describe a U(1) spin liquid\cite{AffleckMarston1988}.   Although these descriptions have been extraordinarily influential, they are uncontrolled, and within  mean-field theory are often unstable  to valence bond ground-states\cite{ReadSachdev1990}. By contrast,  Kitaev's Majorana approach to spin liquids\cite{kitaev,TrebstHickey2022} yields stable $\mathbb{Z}_2$ spin liquids, providing a way to place the FL$^*$ concept on a firmer mathematical  foundation. 
In  previous work\cite{Previous} three of us outlined a microscopic model of an FL$^*$\cite{CPT1} consisting of a three-dimensional Kondo lattice model where the spin subsystem constituted a spin liquid with a Majorana Fermi surface. In this paper we move closer to the cuprates and describe a generalization of this model to a square lattice.

\begin{figure}[h]
    \centering
    \includegraphics[width=1\linewidth]{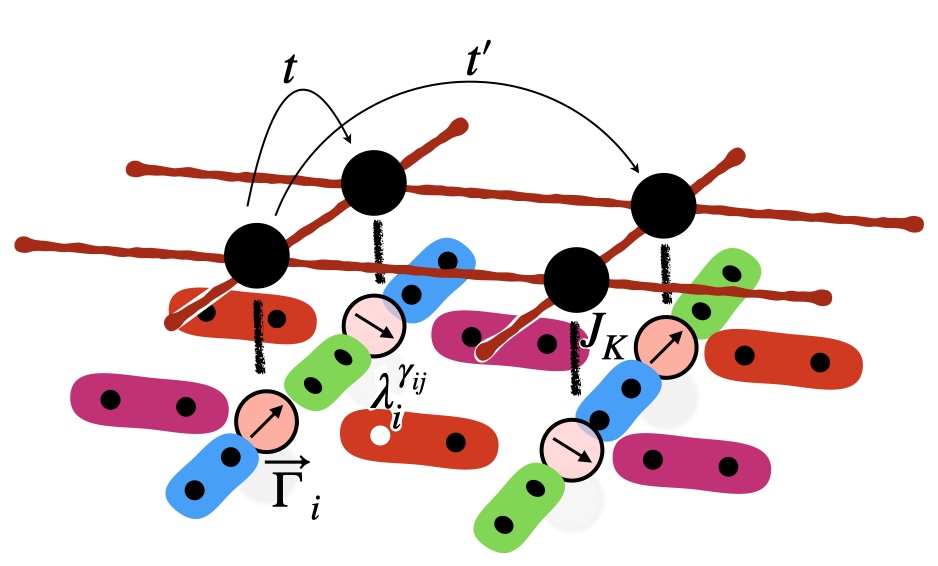}
    \caption{
    Schematic figure of the model. The top layer represents a conduction sea, where conduction electrons reside on a square lattice and hop between nearest neighbors with amplitude $t$ and between next-nearest neighbors with amplitude $t'$. The bottom layer is the generalized Yao--Lee spin liquid~\cite{YL, Vojta2} on a square lattice, consisting of seven Clifford operators. Four of these, $\lambda_i^\gamma$, participate in Kitaev-like anisotropic frustrated interactions $\lambda_i^{\gamma_{ij}}\lambda_j^{\gamma_{ij}}$, with bond-dependent couplings labeled red ($\gamma=1$), blue ($\gamma=2$), purple ($\gamma=3$), and green ($\gamma=4$). The remaining three Cliffords  form spin degrees of freedom $\boldsymbol{\Gamma}_i$ and maintain an $SU(2)$-symmetric representation when projected to a subspace (see explanations in main text) . }
    \label{fig:model}
\end{figure}

Our approach is guided by the following logic. We aim to test whether coupling electrons to a hidden spin liquid can reproduce aspects of the cuprate phenomenology.  Instead of attempting a microscopic derivation, we adopt the simplest analytically tractable model compatible with the square lattice. 
 To avoid the difficulties inherent to typical spin liquid theories  we adopt a Kitaev approach, describing  an integrable two dimensional spin-liquid
which reduces to free fermions. We demonstrate that the 
resulting model 
captures aspects of the phenomenological Yang--Rice--Zhang (YRZ) Green's function\cite{YRZ}, together with a quantum critical point with a logarithmically divergent Sommerfeld coefficient~\cite{zommerfeld}. {While the main subject of study are experimentally accessible, finite temperature observables, we also briefly comment on the underlying zero temperature quantum phase diagram.}

This manuscript is organized as follows: In Sec.~\ref{sec:model} we introduce the model under consideration. It features a weak-coupling Fermi surface instability which we study on the mean-field level in Sec.~\ref{sec:MFT}. The associated coherence factors and experimentally measurable spectral weight are presented in Sec.~\ref{sec:GF}. The impact of quantum and thermal fluctuations are discussed in Sec.~\ref{sec:Fluct}. We conclude with a discussion and outlook and relegate technical details to two appendices.

\section{The model}
\label{sec:model}

Our goal is to construct an SU(2)-symmetric Kondo-lattice model in which conduction electrons are coupled to a stable spin liquid. The spin liquid should host fermionic excitations with a Fermi surface and thus accommodate part of the Luttinger volume. As in the ancilla construction~\cite{ALM1,sachdev25}, the model admits two phases: a large Fermi surface phase in which the electrons and spin liquid are decoupled, and a small Fermi surface phase in which the spins are incorporated into a common Fermi sea.

The only known tractable spin liquid model on a square lattice with the properties required for this purpose is a generalized Yao–Lee spin–orbital liquid~\cite{YL,Vojta2}. { It provides an exactly solvable square-lattice realization of Anderson’s RVB paradigm, with a nonmagnetic spin sector, fractionalized fermionic excitations, an emergent $\mathbb{Z}_2$ gauge field, and} a spinon Fermi surface. We Kondo-couple it to conduction electrons. The Hamiltonian $H = H_{\rm c}+H_{\rm YL}+H_{\rm K}$ can be decomposed as
\begin{subequations}\label{eq:MainModel}
\begin{align}
H_{\rm c} &= -\sum_{i,j}
t_{ij}\bigl(c^\dagger_{i,\sigma}c_{j,\sigma} + \mathrm{H.c.}\bigr), \\
H_{\rm YL} &= -\frac{K}{2}\sum_{\langle i,j\rangle}(\lambda^{\gamma_{ij}}_i\lambda^{\gamma_{ij}}_j)
\Bigl[1- \bm{\Gamma}_i\cdot \bm{\Gamma}_j\Bigr] - h\sum_{\square} B_{\square}, \label{eq:YLKappaTauRho} \\
H_{\rm K} &= \frac{J}{2}\sum_i (c^\dagger \boldsymbol{\sigma} c)_i \cdot (1+\kappa)\boldsymbol{\Gamma}_i .
\end{align}
\end{subequations}
These three terms define, respectively, a conduction-electron band $H_{\rm c}$, a generalized Yao--Lee spin--orbital liquid $H_{\rm YL}$, and a local Kondo coupling $H_{\rm K}$ between them. We describe their meaning, operator content, { and fermionization} below.

\emph{Conduction electrons.} $H_{\rm c}$ describes spin-$1/2$ electrons, where $c_{i\sigma}$ ($\sigma=\uparrow,\downarrow$) annihilates an electron on site $i$, and $t_{ij}$ denotes the nearest- and next-nearest-neighbor hopping amplitudes on the square lattice. Here ``H.c.'' denotes the Hermitian conjugate, and repeated spin indices are summed unless stated otherwise.

\emph{Spin liquid.} $H_{\rm YL}$ describes a generalized Yao--Lee spin--orbital liquid~\cite{Vojta2}. The local Hilbert space on each site is eight-dimensional and is spanned by the identity and seven mutually anticommuting Clifford matrices: four orbital operators $\lambda_i^\gamma$ ($\gamma=1,\ldots,4$) and three spin operators $\boldsymbol{\Gamma}_i=(\Gamma_i^x,\Gamma_i^y,\Gamma_i^z)$,
\begin{subequations}
\begin{align}
\{\lambda_i^\gamma,\lambda_i^{\gamma'}\} & = 2\delta^{\gamma\gamma'}, \quad \{\Gamma_i^{\sigma}, \Gamma_i^{\sigma'} \}  = 2\delta^{\sigma\sigma'},\\
\{\lambda_i^\gamma,\Gamma_i^{\sigma}\}& = 0, \qquad {\sigma\in\{x,y,z\}.}
\end{align}
\end{subequations}

All matrices at a given site anticommute, whereas matrices on different sites commute. A key feature of the model is the bond variable $\gamma_{ij}\in{1,2,3,4}$, denoted by red, blue, purple, and green bonds in Fig.~\ref{fig:model}, which links the $A$ and $B$ sublattices. 

The interaction term proportional to $K$ acts on nearest-neighbor bonds $\langle i,j\rangle$. The bond label $\gamma_{ij}\in{1,2,3,4}$ selects the corresponding orbital component $\lambda^{\gamma_{ij}}$ entering the coupling. The orbital sector thus enters through the bond-dependent factor $\lambda_i^{\gamma_{ij}}\lambda_j^{\gamma_{ij}}$, while the bracketed term couples the spin sector via $\bm{\Gamma}i\cdot\bm{\Gamma}j$. 
The plaquette term proportional to $h$ is an ordered product of bond operators around $\partial\square$, $B_{\square} =\prod_{(ij)\in\partial\square}\lambda^{\gamma_{ij}}_{i}\lambda^{\gamma_{ij}}_{j}$. { As established in Ref.~\cite{Vojta2}, this model has an $SO(4)=SU(2)\times SU(2)$ symmetry, which will play an important role in the physics discussed below.}

{ Like the Kitaev spin liquid, this generalized Yao--Lee spin--orbital liquid is exactly solvable by fermionization~\cite{kitaev,Vojta2}. Following Ref.~\cite{Vojta2}, we represent the seven Clifford operators as bilinears of eight Majorana fermions $b_i^\alpha$ ($\alpha=1,\ldots,8$, with $\{b_i^\alpha,b_i^\beta\}=2\delta^{\alpha\beta}$),
\begin{align}
\lambda_i^\gamma = \ri\, b_i^8 b_i^\gamma \quad (\gamma=1,\ldots,4),
\\ {\bf\Gamma}_i^a = \ri\, b_i^8 b_i^{a+4} \quad (a=1,2,3),
\label{eq:MajRep}
\end{align}
subject to the on-site constraint $b_i^1\cdots b_i^8=-1$. Under this substitution, the bond interaction depends on the four orbital Majoranas $b^{1,\ldots,4}$ only through the bilinear $\hat u_{(i,j)} = \ri\, b_i^{\gamma_{ij}} b_j^{\gamma_{ij}}$, and $H_{\rm YL}$ becomes quadratic in the four remaining Majoranas $b^{5,\ldots,8}$,
\begin{equation}
H_{\rm YL} = \frac{\ri}{2}K\sum_{\langle i,j\rangle}\sum_{\alpha=5}^{8}\hat u_{(i,j)}\, b_i^\alpha b_j^\alpha - h\sum_{\square} B_{\square}.
\label{eq:HYLmaj}
\end{equation}
Since $\hat u_{(i,j)}$ commutes with $H_{\rm YL}$ and squares to unity, it is a static $\mathbb{Z}_2$ gauge field taking values $\hat u_{(i,j)}=\pm 1$. The matter sector is therefore a system of noninteracting Majorana fermions in the background of this static gauge field. The four matter Majoranas on each site combine into two complex fermions,
\begin{equation}
f_{i\uparrow} = \tfrac{1}{2}\bigl(b_i^5 + \ri\, b_i^6\bigr),
\qquad
f_{i\downarrow} = \tfrac{1}{2}\bigl(b_i^7 + \ri\, b_i^8\bigr),
\label{eq:fdef}
\end{equation}
up to a sublattice-dependent phase; see Appendix~\ref{app:YLDetails}. We identify these as the two spin components $\sigma=\uparrow,\downarrow$ of a single complex spinon $f_{i\sigma}$. In terms of these fermions, the spin liquid becomes a spinon hopping problem,
\begin{equation}
H_{\rm YL} = -K\sum_{\langle i,j\rangle}\hat u_{(i,j)}\bigl(f^\dagger_{i\sigma} f_{j\sigma} + \mathrm{H.c.}\bigr) - h\sum_{\square} B_{\square}.
\label{eq:HYLf}
\end{equation}
The flux term proportional to $h>0$, which we assume to be the largest energy scale in the problem, stabilizes the uniform-flux sector, whose ground state carries a spinon Fermi surface.}

\emph{Kondo coupling.} $H_{\rm K}$ couples the local spin density of the conduction electrons, $(c^\dagger \boldsymbol{\sigma} c)_i$, to the spin-liquid operator $\frac{(1 + \kappa)}{2}\boldsymbol{\Gamma}_i$. Here $\kappa=-\ri\,\Gamma^5\Gamma^6\Gamma^7 = \lambda^1 \lambda^2 \lambda^3 \lambda^4$. Within the $\kappa=1$ subspace, the projector $\frac{(1+\kappa)}{2}$ filters the spin components of the ancillary fluid so that $\frac{(1 + \kappa)}{2}\boldsymbol{\Gamma}$ satisfies the SU(2) commutation algebra. {In the Majorana representation above, this operator is precisely the spinon spin density,
\begin{equation}
\frac{(1+\kappa)}{2}\boldsymbol{\Gamma}_i = (f^\dagger\boldsymbol{\sigma} f)_i,
\end{equation}
so that $H_{\rm K}$ couples the conduction-electron spin to the spinon spin.}

\emph{Fermionized Hamiltonian.} { Collecting these terms, the full model becomes a fermionized $\mathbb{Z}_2$ Kondo lattice,}
\begin{align}
H[c,f] ={}& -\sum_{i,j} t_{ij}\bigl(c^\dagger_{i\sigma} c_{j\sigma} + \mathrm{H.c.}\bigr)\notag
\\&- K\sum_{\langle i,j\rangle} \hat u_{(i,j)}\bigl(f^\dagger_{i\sigma} f_{j\sigma} + \mathrm{H.c.}\bigr)
- h\sum_{\square} B_{\square}\notag
\\&+ J\sum_i (c^\dagger \boldsymbol{\sigma} c)_i \cdot (f^\dagger \boldsymbol{\sigma} f)_i,
\label{HF}
\end{align}
{ where the flux is now expressed through the gauge fields,} $B_{\square} = \prod_{(ij)\in\partial\square} \hat u_{(i,j)}$. We use the convention that $(i,j)$ orders the indices so that the first index lies on the $A$ sublattice; with this convention, $\hat u_{(i,j)} = \hat u_{(j,i)} = \pm 1$ is symmetric. {{Further} details are given in Appendix~\ref{app:YLDetails}.}

Equation~\eqref{HF} resembles a conventional Kondo lattice, but with two important distinctions. First, there is no local constraint imposed on the Hilbert space of the $f$ fermions, which makes the model tractable using conventional many-body techniques. Second, there is a discrete $\mathbb{Z}_2$ gauge symmetry,

\begin{eqnarray}
    f_{j\sigma}&\rightarrow &s_jf_{j\sigma},\cr
    u_{(i,j)}& \rightarrow & s_i u_{(i,j)}s_j, \qquad (s_i, s_j = \pm 1).
\end{eqnarray}
{ The fermionic representation also makes manifest the global $SU(2)\times SU(2)$ symmetry: one $SU(2)$ corresponds to rotations of the spin indices, while the other acts in particle-hole space,
\begin{equation}
f_{j,\sigma}\rightarrow \epsilon_{\sigma\sigma'}f^\dagger_{j,\sigma'} .
\end{equation}
}

\section{Mean field (RPA) treatment}
\label{sec:MFT}

We treat the Kondo interaction in Eq.~(\ref{HF}) at the saddle-point (RPA) level using a Hubbard--Stratonovich transformation. This is motivated by leading diagrams with large logarithms.

 To this end, we first technically express the theory using a many-body path integral
\begin{subequations}
\begin{align}
    \mathcal Z &= \int \mathcal Dc \mathcal D f \; e^{- S} \\
    S & = \int_0^\beta d\tau \sum_{i, \sigma}\bar c_{i, \sigma} \partial_\tau c_{i, \sigma} + \bar f_{i, \sigma} \partial_\tau f_{i, \sigma}  + H[c,f], \label{eq:SOriginal}
\end{align}
\label{eq:PathInt}
\end{subequations}
where $c_{i,\sigma}, f_{i, \sigma}$ are now Grassmann fields.

\subsection{Channels of Fermi surface instabilities}

The decoupling is organized using the Sp(2) completeness relation\cite{flintcoleman2009} 
\be
\sum_{a=1}^3 (\sigma^a)_{\alpha\beta}(\sigma^a)_{\gamma\eta}
= \delta_{\alpha\eta}\delta_{\beta\gamma}-\epsilon_{\alpha\gamma}\epsilon_{\beta\eta}.
\label{eq:FierzPauli}
\ee
The local exchange can be rewritten in terms of SU(2)-singlet particle--hole (hybridization) and particle--particle (pairing) bilinears,

\be
\hat V_i \equiv f^+_{i\s}c_{i\s}(-1)^{x+y},\qquad
\hat\Delta_i \equiv c^+_{i\s}\epsilon_{\s\s'}f^+_{i\s'}.
\ee

We decouple the hybridization and pairing channels using Hubbard--Stratonovich fields $V_i$ and $\Delta_i$ respectively. Using this procedure the path integral becomes
\begin{subequations}
    \begin{align}
        \mathcal Z &= \int \mathcal Dc \mathcal D f \mathcal DV \mathcal D\Delta\; e^{-S_{\rm HS}} \\
    S_{\rm HS} & = S\vert_{J = 0} + \int_0^\beta d\tau \sum_i \Big[\frac{|V_i|^2+|\Delta_i|^2}{J} \notag\\
&+\bigl(V_i\,c^+_{i\s}f_{i\s}(-1)^{x+y}+\mathrm{H.c.}\bigr)
+\bigl(\Delta_i\,c^+_{i\s}\epsilon_{\s\s'}f^+_{i\s'}+\mathrm{H.c.}\bigr)
\Big],
\label{eq:HSdec}
    \end{align}
\end{subequations}
with $S\vert_{J =0}$ is Eq.~\eqref{HF} evaluated at $J = 0$.

At the saddle point this leads to staggered hybridization and onsite pairing,
\be
V = 
\la c^+_{i\s}f_{i\s}\ra(-1)^{x+y},
\qquad
\Delta = \la c^+_{i\s}\epsilon_{\s\s'}f^+_{i\s'}\ra. 
\ee

\subsection{Mean-field Hamiltonian}

The gauge choice $\hat u_{(i,j)} = 1$ in Eq.~\eqref{HF} aligns the spinon dispersion with the conduction band, yielding { the following expressions for the conduction and f-electron dispersion}:
\bea
\label{disp}
&& \epsilon_{k} = - 2t(\cos k_x +\cos k_y) +4t'\cos k_x\cos k_y-\mu, \label{t}\cr
&& K_{k} = -{2}K(\cos k_x +\cos k_y). 
\eea 
To write the Green's functions compactly, we introduce the Nambu spinor (with ${\bf Q}=(\pi,\pi)$)
\be
\Psi_k \equiv \begin{pmatrix}
 c_{k, \uparrow} &
c^\dagger_{-(k+Q),\downarrow} &
 f_{k+Q, \uparrow} &
f^\dagger_{-k, \downarrow}
\end{pmatrix}^T,\label{eq:Nambu4spinor}
\ee

This choice mixes $c_k$ with $f_{k+Q}$ in the particle block and $c^\dagger_{-(k+Q)}$ with $f^\dagger_{-k}$ in the hole block.  The mean field Hamiltonian  then consists of two independent blocks according to spin. 

The mean-field action then reads 
\be
S_{\rm MF}= \sum_{k,\omega_n}\Psi_{k,\omega_n}^+\,(-\ri \omega_n+h_k)\,\Psi_{k,\omega_n}
+\frac{\beta}{J}\bigl(|V|^2+|\Delta|^2\bigr),
\ee
with with mean field Hamiltonian consisting of two identical blocks corresponding to different spin projections 
\be
h_k=
\left(
\begin{array}{cccc} \epsilon_k & 0 & V & \Delta\\
0 & -\epsilon_{-(k+Q)} & \Delta^\ast & -V^\ast\\
V^\ast & \Delta &K_{k+Q} & 0\\
\Delta^\ast & -V & 0 & -K_{k}
\end{array}
\right).
\label{eq:NambuGinv}
\ee

The order parameter matrix is 
\begin{equation}
   {\cal V} = \left ( \begin{array}{cc}
        V & \Delta \\
        \Delta^* & - V^* 
    \end{array} \right ) = i\sqrt{\vert V \vert^2 + \vert \Delta \vert^2} U, \label{U}
\end{equation}
 where $U$ is an SU(2) matrix. { The $SU(2)$ symmetry of the order parameter reflects the underlying particle-hole symmetry of the spin liquid.} This matrix can be diagonilized by the unitary transformation of the $f$-operators resulting in the electron dispersion Eq.~\eqref{B9} (see the Appendix~\ref{app:GLDetails}). 

 \subsection{Free energy}

 When we integrate out the fermions, symmetry considerations dictate the following form of the effective free energy functional:
 \begin{align}
\label{free}
{\cal F} &= -\rho
\sum_{\langle ij\rangle} 
u_{(i,j)} \mbox{Tr}({\cal V}_j{\cal V}^\dagger_i + 
{\cal V}_i {\cal V}^\dagger_j) 
\cr
&\quad + \sum_i \Big[
\tau {\rm Tr}({\cal V}_i^\dagger {\cal V}_i)
+ g_1 {\rm Tr}({\cal V^\dagger}_i{\cal V}_i)^2
\Big] \cr
&- \kappa \sum_{\square}\prod_\square u_{(i,j)} .
\end{align}

The square plaquette terms determine the vison energy. The  uniform part of the free energy  depends on  ${\cal V}{\cal V^\dagger} =(|V|^2+|\Delta|^2)I$. We remark that, as in all Landau-Ginzburg theories, the  free energy functional implicitly assumes slow variations of fields on the scale of the lattice constant (modulo discrete gauge transformations in the present case).

\begin{figure*}
        \centering
    \includegraphics[width=\textwidth]{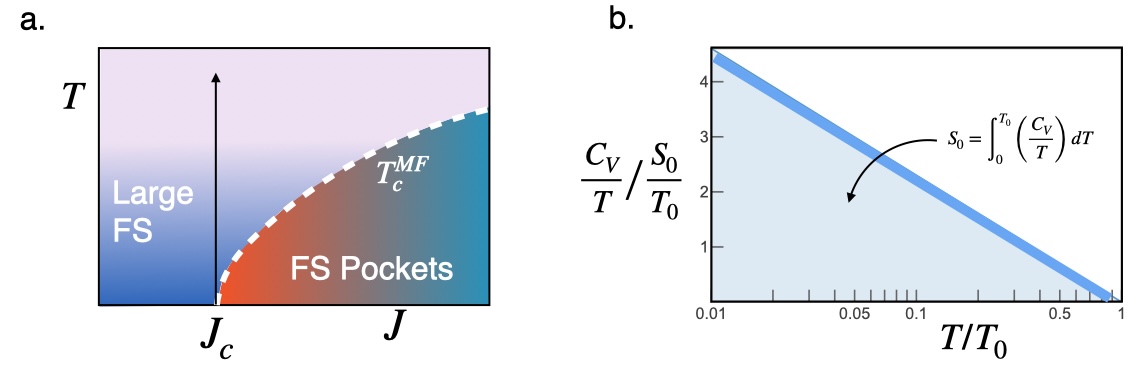}
    \caption{ {{\bf a.} Phase diagram showing the mean-field transition from large Fermi surface to small Fermi pockets as a function of Kondo coupling $J$ between the conduction electrons and ancillary spins. $T_c^{MF}$ marks a mean-field crossover scale. {\bf b.} Logarithmic  ``Sereni" dependence of the Sommerfeld coefficient on temperature, evaluated at the critical exchange coupling $J_c$.}} %along the dotted line trajectory shown in {\bf a.}
    \label{fig:schematic}
\end{figure*}

The parameter $\tau$ undergoes a sign change signaling the emergence of a finite amplitude of ${\cal V}$ and thus a mean-field phase boundary in $T$ vs $\mu$ space. 
Specifically (see Appendix~\ref{app:GLDetails}), we find $\tau \sim (T/T_c-1)/J$ with {$T_c= 16\sqrt{Kt} e^{- \sqrt{\frac{\pi^2 (t + K)}{4J}+\ln^2 4-  \ln^2 \left ( \sqrt{\frac{t}{K}}\right )}}$} to  leading (zeroth) order in $t', \mu$ \eqref{mfTc}. By contrast, at zero temperature and to zeroth order in $t'$ we find $\tau \sim (\vert \mu \vert/\mu_c-1)/J$ where  {$\mu_c$ is given by Eq.(\ref{muc}), in particular for $t \gg K$ we find $\mu_c \simeq T_c$. Figure \ref{fig:schematic}{\bf a.} illustrates the mean-field phase diagram. }

To incorporate phase fluctuations {(see Sec.~\ref{sec:Fluct})} below the crossover temperature, we retain the leading gradient terms as in (\ref{free}). Away from the quantum critical point the amplitude of ${\cal V}$ remains fixed.

\section{Green's function, coherence factors and spectral weight}
\label{sec:GF}

\begin{figure}[h]
    \centering
    \includegraphics[width=1\linewidth]{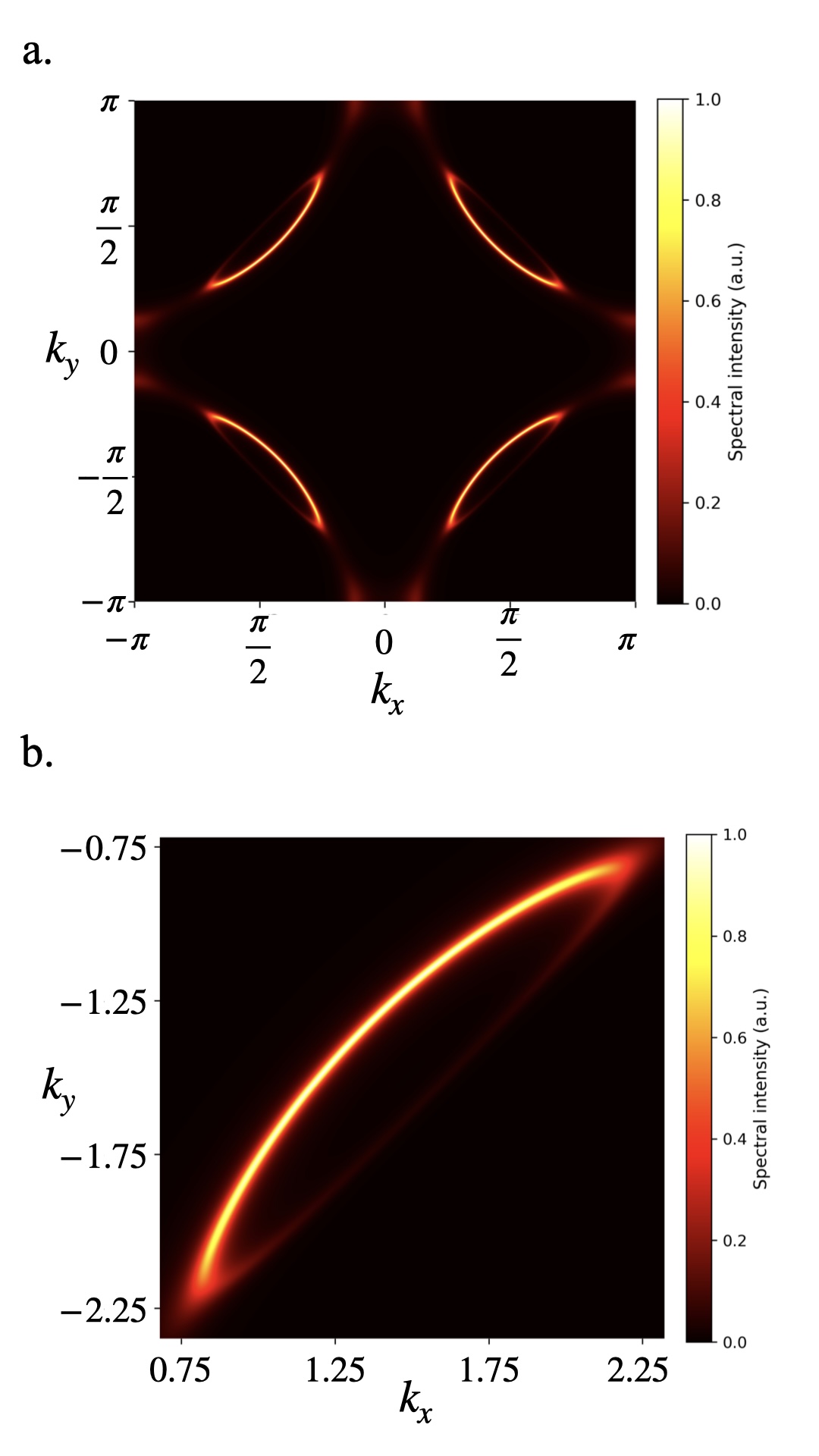}
    \caption{Simulated ARPES intensity from the YRZ spectral function, Eq.~\eqref{YRZ}, for $t=0.5$, $t'=-0.159$, $K=0.32$, $\mu=-0.44$, and $V=0.08$.  (a) The spectral intensity exhibits four Fermi arcs characteristic of the pseudogap phase. (b) Magnified view of the lower-right Fermi arc, revealing that it is part of a closed Fermi pocket. The weak backside of the pocket has predominantly spinon character due to Kondo hybridization and is associated with the transition from a large to a small Fermi surface in the pseudogap phase. }
    \label{fig:ARPESFermiArc}
\end{figure}

In this section we discuss the Green's function $G(z,k) = [z - h_k]^{-1}$
as obtained using the mean-field Hamiltonian, Eq.~\eqref{eq:NambuGinv}. Since the $f$-fermion propagator is trivial in Nambu space, it follows that the $c-c$ correlator does not contain anomalous (Gor'kov-like) off-diagonal components. Thus, even at finite $\Delta$, the system is not a superconductor.

The only non-trivial component of the conduction electron Green's function follows from Eq.~\eqref{t} together with Eq.~\eqref{eq:NambuGinv}, 

\bea
 G_{{cc}}(z,k) = \Big[z - \epsilon_{k} - \frac{\vert V \vert^2 + \vert \Delta \vert^2}{z + K_k }
 \Big]^{-1}. \label{YRZ}
 \eea

Equation~\eqref{YRZ} is the Yang--Rice--Zhang (YRZ) form\cite{YRZ}, { which has been successfully} used as a  phenomenological parametrization of the cuprate pseudogap and commonly fit to ARPES spectra {\cite{Peter}, to analyze autocorrelation of photoemission data \cite{bascones2008} and most recently to reproduce the observations of the Yamaji effect and quantum oscillations \cite{Yamaji2025}}. Importantly, in the context of model~(\ref{eq:MainModel}), it follows in a controlled fashion considering the leading Fermi-surface instability at the mean-field level {and the self-energy term emerges from coherent scattering between $c$ and $f$ electrons.}.

\begin{figure}[h]
 \includegraphics[width=1\linewidth]{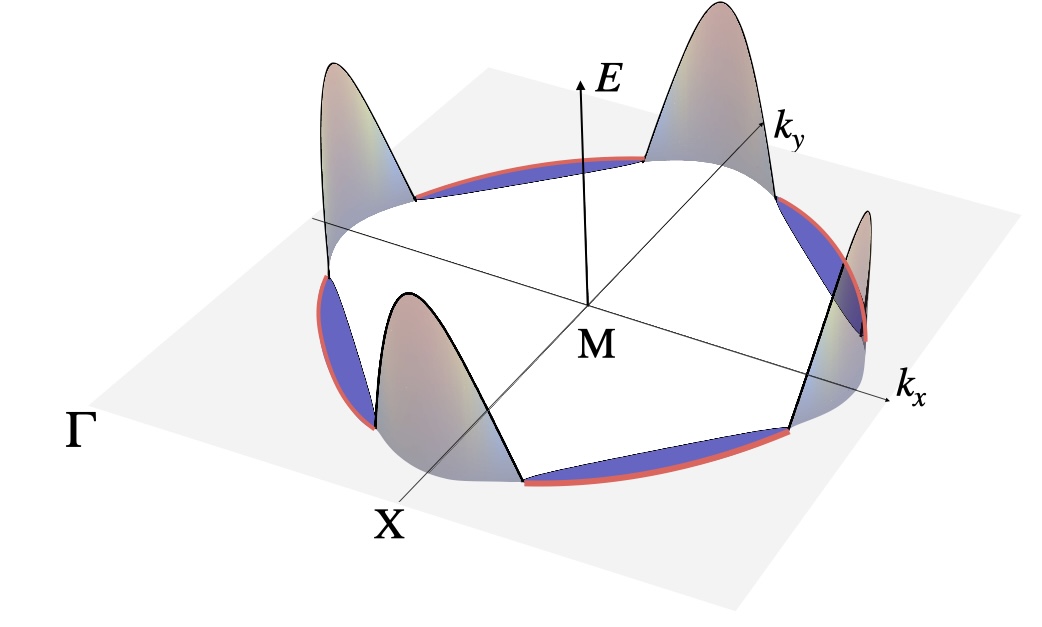} 
 \caption{Showing the gap profile around the reconstructed Fermi pockets centered at the M point. The blue regions denote the reconstructed Fermi pockets, where the red curve denotes the electron edge of the pocket. The vertical direction shows the size of the gap in regions between the pockets. The Figure was constructed from a Mathematica plot with parameters $t=0.5$, $t'=0.13$, $K_0=0.64$, $\mu=-0.44$, $V_0=0.08$. }
 \label{fig:TwoBishops}
\end{figure}

In addition to poles signaling the reconstructed (small) Fermi surface, $G_{{cc}}(z,k)$ also displays Green's function zeros~\cite{AGD,Dzyaloshinskii2003} at the location of the original spinon Fermi surface\cite{Fabrizio2023}. Moreover the spectral weight $- \text{Im}[G_{{cc}}(\omega+i0, k)]/\pi$, which is experimentally measurable using ARPES (Fig. \ref{fig:ARPESFermiArc}), displays strongly $k$-dependent coherence factors determining the relative weight of $c$ and $f$ electrons for eigenstates $\vert{\psi_k} \rangle$ at a given momentum. 
\begin{equation}
    \vert{\psi_k} \rangle = \cos(\theta_k/2) \vert{c_k}\rangle + \sin(\theta_k/2) \vert f_k \rangle,
\end{equation}
where $\tan(\theta_k) = 2[\vert V \vert^2 + \vert \Delta \vert^2]/[\epsilon_k + K_k]$. 
The spectral weight is small in the vicinity of the antiferromagnetic Brillouin zone so that the Fermi-pockets resemble arcs rather then being closed (Figs. \ref{fig:ARPESFermiArc},\ref{fig:TwoBishops}).

\section{Fluctuations}
\label{sec:Fluct}

Building on the previous discussions of the mean-field solution, we here discuss the role of quantum and thermal fluctuations beyond the mean-field approximation.

\subsection{Gauge fluctuations and Elitzur's theorem}

 The Hubbard-Stratonovich fields inherit an emergent $\mathbb Z_2$ gauge charge from their constituent $f$ fermions. Under the  gauge transformation $f_i \rightarrow s_i f_i, \hat u_{ij} \rightarrow s_i \hat u_{ij} s_j$, these fields transform as  $ V_i, \Delta_i \rightarrow s_i V_i, s_i\Delta_i$, $s_i = \pm 1$.

 Our microscopic model, Eq.~\eqref{eq:MainModel} corresponds to Eq.~\eqref{HF}, a static gauge theory (i.e. all plaquette operators $B_\square$ are conserved). In this limit, we perform gauge fixed calculations and find ``spontaneous symmetry breaking'' and associated ``order parameters'' ($V, \Delta$) { Recall that in two dimensions the order can occur only at $T=0$ and we can safely postpone the consideration of this problem, see the discussion in subsection V.B.} In a more general setting without gauge fixing, the correlators displaying long-range orders are of the type $V_i^* S_{ij} V_j$ where $S_{ij}$ is a gauge string (a product of $\hat u_{ij}$s) connecting lattice sites $i,j$ along an arbitrary contour. Crucially, the combination $V_i^* S_{ij} V_j$ is itself gauge invariant {, physically measure the coherence of $c-$to-$f-$conversion,} and can display long-range order while expectation values of $V_i^* V_j$ vanish by cycling over gauge configurations (Elitzur's theorem).

 { In the static gauge limit, Eq. (23) should be understood as the gauge-fixed expression for the electron Green’s function in a fixed flux sector. Gauge-invariant coherence is measured by string-dressed $c$-to-$f$ conversion correlators. Once finite temperature fluctuations are included, true long-range
order is destroyed, but the mean-field Green’s function remains a controlled description below the correlation length relevant for finite-energy probes such as ARPES and magnetotransport. } 

Without going into technical details, we briefly discuss our expectations beyond the static limit of the gauge theory (i.e. we briefly include ``electric terms'' of strength $g$ into Eq.~\eqref{HF}, as induced, e.g., by non-Yao-Lee spin-orbital interactions). Adding small $J$, we expect that the fractionalized, deconfining phase of the spin liquid to survive, as it is protected by the vison gap $\mathcal O(K)$. This however implies a finite string tension and thus $ \langle V_i^* S_{ij} V_j \rangle \sim e^{-\vert i - j \vert/\xi}, \xi \sim 1/g$. Thus, quantum fluctuations in the present 2D model are expected to have a similar effect as thermal fluctuations in a related 3D, $\mathbb Z_2$ gauged $XY$-model studied by two of us~\cite{Kuklov} {and restrict the validity of the mean-field expression Eq.~\eqref{YRZ} to subexponential length scales (note that both ARPES and magnetotransport probe the system at finite length scales).} At larger $J$ we expect a quantum phase transition into a Higgs (i.e. non-fractionalized) phase, but leave details of these  questions to future studies.

In the following we return to the simpler situation of the static $\mathbb Z_2$ gauge theory. For simplicity we slightly stretch the standard nomenclature and use words such as ``symmetry breaking, Mermin-Wagner theorem'' etc.

\subsection{ Away from the critical point. Thermal fluctuations}

 Below the mean field transition temperature and far from the quantum critical point the order parameter amplitude is stabilized. 
 The low-energy fluctuations are those of the SU(2) matrix $U$ defined by Eq.(\ref{U}). 
 The resulting  SU(2) nonlinear sigma model (NLSM) describes thermal fluctuations deep in the pseudogap state: 
\bea
F = -\frac{\rho}{2}\sum_{<i,j>}u_{i,j}\mbox{Tr} (U^\dagger_i U_j +H.c.)- \kappa \prod_{\square} u_{(ij)}, \label{PCF}
\eea
 
Mermin-Wagner arguments tell us that  the $U_{i}$ fields will strongly fluctuates so that  true long-range order of $\mathcal V$ can not be established at finite temperatures. { The thermal fluctuations destroy the long range order even when fluctuations of the gauge field are not taken into account.  
(we will show in Appendix C that for $\kappa <<T$ they are irrelevant).
Then the continuum limit of model (\ref{PCF}) becomes the exactly solvable Principal Chiral Field model \cite{Wiegmann1984} for which it was firmly established that the correlation length is finite: $\xi(T) \sim \exp(2\pi\rho/T)$. } Another source of disorder are thermally excited visons. Since $U_{i}$ carries $\mathbb Z_2$ charge, one must include the $\mathbb Z_2$ gauge field into the low energy description, which makes a continuum limit formulation difficult.  
%\textcolor{magenta}{[EJK: I would leave the next bit out out and reserve it for our future publication.]}
%\textcolor{red}{[PC: I agree.]}
%{ We will show, as promised, that these fluctuations do not modify the correlation length. For this end we chose a particular gauge for the $u_{ij}$ field setting its value to 1 except for $(na_x, (n+1)a_x; 2ma_y, 2ma_y$ links where it can be either $+1$ or $-1$. This way  every plaquette can have flux $\pm 1$ and the re is no over-summation. Summing on $u_{ij}$ on these links and expanding in $\exp(-\kappa/T)$ we obtain the following contribution to the partition function:
%\bea
%&& Z_{(x,2y);(x+a,2y)} = \\
%&& \exp\Big[ \frac{\rho}{2T}\mbox{Tr}(U_{x,2y}U\dg_{x+a,2y}+ H.c)\Big] +\nonumber\\
%&& \re^{-\kappa/T}\exp\Big[ -\frac{\rho}{2T}\mbox{Tr}(U_{x,2y}U\dg_{x+a,2y}+ H.c.)\Big] \approx \nonumber\\
%&& \exp\Big\{ -\frac{\rho a^2}{2T}\mbox{Tr}(\p_x U\p_x U\dg)\Big[1 +2\exp[(-\kappa-2\rho)/T)]\Big]\Big\} .\nonumber
%\eea
%So the vison contribution in this limit is reduced to a renomalization of stiffness.  In the continuum limit the model becomes the PCF one. 
%}

We note in passing that this is the same sigma model as in 
Refs.~\cite{CPT1,CPT2,CPT3}, but  in one dimension lower. 

 Besides the $\mathbb Z_2$ charge field $U$ carries electric charge $e$.  Recall that in contrast to the Kondo lattice model where the phase of $V$ could be eliminated by the gauge transformation of the $f$-electron operator and absorbed in the Lagrange multiplier field, this cannot be done here. So, $V$ and $\Delta$ are genuine charge $e$ fields.  So, below the phase transition we must expect Meissner effect. However, as we have said, in 2D non-Abelian sigma models such as (\ref{PCF}) the thermal fluctuations transform the transition into a crossover. Hence instead of the Meissner effect we expect strong diamagnetic fluctuations. 

\subsection{Fluctuations near the Quantum Critical Point}

 { The perfect nesting between the quadratic spinon Fermi surface and the Fermi surface of the conduction electrons is only present at $\mu =0, ~~t'=0$.} Away from perfect nesting, the particle--hole bubble is finite, so the hybridization instability occurs only beyond a finite critical coupling $J_c(p)$ (with $p$ the doping, see $\mu_c$ introduced above). Accordingly, the model~(\ref{HF}) supports two regimes: a decoupled phase with a large electron Fermi surface and a hybridized phase with a small Fermi surface, as displayed in Figs.~\ref{fig:ARPESFermiArc}, \ref{fig:TwoBishops}. In this Section we study signatures of quantum fluctuations in thermodynamics and transport, as the quantum critical point is approached from $J < J_c(p)$.

Within a large-$N$ extension (rendering the RPA controlled), the hybridization fluctuations are described by the RPA bosonic propagator,
\be
D(\Omega,q) \equiv \langle {\cal V}(\Omega,q){\cal V}^\dagger(\Omega,q)\rangle
= \Big[J^{-1}-\Pi_{cf}(\Omega,q)\Big]^{-1}, 
\ee
where $\Pi_{cf}$ is the electron--spinon polarization bubble,
\bea
&& \small\Pi_{cf}(\ri\Omega_m,q) = \label{eq:Pi_cf}\\
&& 2T\sum_{\omega_n}\int \frac{\rd^2 k}{(2\pi)^2}
\,G_{cc}(\ri\Omega_m+\ri\omega_n,k)\,G_{ff}(\ri\omega_n,k+{\bf Q}+q),
\nonumber
\eea
with ${\bf Q}=(\pi,\pi)$. The low-energy dynamics of the bosonic propagator is controlled by the ``hot spots'' $k_h$ defined by the intersection of the two Fermi surfaces after translation by ${\bf Q}$,
\be
\epsilon(k_h)=0,\qquad K(k_h+{\bf Q})=0.
\ee
Linearizing near $k_h$ and performing the analytic continuation $\Omega_m\to \Omega+\ri 0^+$ yields a Landau-damped polarization, 
\begin{align}
\Im m\,\Pi_{cf}^{(R)}(\Omega,q) &= \gamma\,  \Omega + \cdots, \\
Z^{-1} \equiv\gamma = \frac{2}{\pi}&\sum_{h}\frac{1}{|\vec v_{f,h}\times\vec v_{c,h}|}
= \frac{16}{\pi}\frac{1}{|\vec v_{f,h}\times\vec v_{c,h}|},\label{eq:LandauDamping}
\end{align}
where $\vec v_{c,h}=\nabla_k\epsilon(k)|_{k_h}$ and $\vec v_{f,h}=\nabla_k K(k)|_{k_h+{\bf Q}}$ are, respectively, the conduction-electron and spinon Fermi velocities at the hot spots. For the symmetric square-lattice geometry considered here the sum runs over eight equivalent hot spots, yielding an overall prefactor $8$ when all $|\vec v_{f,h}\times\vec v_{c,h}|$ are equal.

The real (dispersive) part of the polarization bubble is smooth at low energies and can be expanded at small momentum as
\be
\Re e\,\Pi_{cf}^{(R)}(0,q)=\Pi_{cf}(0,0)-Dq^2+\cdots.
\ee
This renormalizes the boson mass and provides the leading $q^2$ stiffness; together with the Landau damping in Eq.~(\ref{eq:LandauDamping}), the retarded RPA propagator takes the diffusive form
\be
D^{(R)}(\Omega,q) = \frac{1}{m + D q^2 + \ri Z^{-1}\Omega},
\label{D}
\ee
with the tuning parameter $m\equiv J^{-1}-\Pi_{cf}(0,0)$ vanishing at the quantum critical point {$J=J_c$}.  The linear frequency term is generated by the Landau damping in the presence of Fermi surface. The suggested effective field theory is { (we take the gauge where $A_0 =0$ ):}
\bea
&& S = \label{actionEM}\\
&&\int \rd \Omega \rd^2 x\, \mbox{Tr}\Big[{\cal V}^+\Big[Z^{-1}|\Omega| + D(\nabla +\ri e {\bf A})^2 +m\Big]{\cal V}\Big] + \cdots.
\nonumber
\eea
Since ${\cal V}$ carries electric charge $e$, its long-wavelength coupling to an external electromagnetic field is obtained by the minimal substitution $\nabla\to\nabla+\ri e\mathbf A$ in Eq.~\eqref{D}. Since $z=2$ the effective dimensionality is $d_{eff} = d+z =4$ and the theory is marginal at the leading RPA level (cf. Refs.~\cite{Millis1993, AbanovSchmalian2003, MetlitskiSachdev2010} for the related spin-fermion problem).

At the Gaussian level, the hybridization fluctuations contribute
\be
\frac{F_{\rm fl}}{V}=\frac{T}{2}\sum_{\Omega_n}\int\frac{\rd^2 q}{(2\pi)^2}\ln D^{-1}(\ri\Omega_n,q),
\ee
with $D$ the bosonic propagator. 
Using the spectral representation of $\ln D^{-1}$ and the retarded response Eq.~(\ref{D}), one can write the entropy in terms of the bosonic phase shift $\delta(\omega,q)=\arg D^{-1}_R(\omega,q)$,
{\small
\begin{align}
\frac{S}{V}
&= \int_{-\infty}^{\infty}\rd\omega\,\frac{\omega}{2T^2\sinh^2(\omega/2T)}
\int \frac{\rd^2 q}{(2\pi)^2}\,\delta(\omega,q),
\\
\delta(\omega,q)
&=\tan^{-1}\Big[\frac{\Im mD^{(R)}(\omega,q)}{\Re e  D^{(R)}(\omega,q)}\Big] =\tan^{-1}\!\left[\frac{Z^{-1}\,\omega}{m+Dq^2}\right].
\end{align}
}
The specific heat follows from $C_v=T\,\frac{\partial}{\partial T}\left(\frac{S}{V}\right)$. For the diffusive propagator in Eq.~(\ref{D}), the $q$ integral is logarithmic in $d=2$, yielding the {``Sereni scaling law'' for the Sommerfeld specific heat coefficient }\cite{Sereni} 
{
\bea
\frac{C_v}{T} = \frac{S_0}{T_0}\ln \left[\frac{T_0}{\mbox{max}(T,Zm)}\right],
\eea
where $T_0 = ZD$ is the UV cutoff of the effective theory (\ref{actionEM}), $S_0=q_0^2\frac{\pi}{6}$ is the integrated entropy $S_0\sim \int_0^{T_0} \frac{C_v}{T} dT$ associated with the critical fluctuations, while $q_0^2$ is of order of the size of the Fermi surface at the quantum critical point (QCP). } This produces a logarithmically enhanced Sommerfeld coefficient (see Fig. \ref{fig:schematic}{\bf b.}), as observed near the cuprate Lifshitz transition\cite{zommerfeld}. Similar results were previously obtained in the context of the spin-fermion model~\cite{Millis1993}, but here  no spin-ordering occurs.

 We further speculate about the implications of the ``order parameter'' fluctuations for transport. As a leading order contribution, we consider the 
corresponding fluctuation (Aslamazov--Larkin) contribution to the conductivity follows from the Kubo formula,
{\small
\begin{align}
\sigma^{\alpha\beta}_{\mathrm{AL}}
&= \frac{2e^2D^2}{\pi }\int \frac{\rd^2k}{(2\pi)^2}\,k_{\alpha}k_{\beta}
\int \frac{\rd\omega}{T\sinh^2(\omega/2T)}\,[\Im m D(\omega, k)]^2
\nonumber\\
&= \frac{e^2}{8\pi^2 Z^2}\,\delta_{\alpha\beta}\int \frac{\rd\omega}{T\sinh^2(\omega/2T)}
\left[1- \frac{Zm}{\omega}\tan^{-1}\!\left(\frac{\omega}{Zm}\right)\right]
\nonumber\\
&\equiv \frac{e^2}{2\pi^2 Z^2}\,\frac{1}{f(Zm/2T)}\,\delta_{\alpha\beta}.
\label{sigma}
\end{align}
}

At the quantum critical point we tune $m(T=0)=0$. The mass acquires a temperature dependence from $m(T)=J^{-1}-\Pi(0,0;T)$, and exhibits a crossover from $m(T)\propto T^2$ at the lowest temperatures to an approximately linear-in-$T$ behavior at higher $T$. 

\begin{figure}[h]
        \includegraphics[width=1\linewidth]{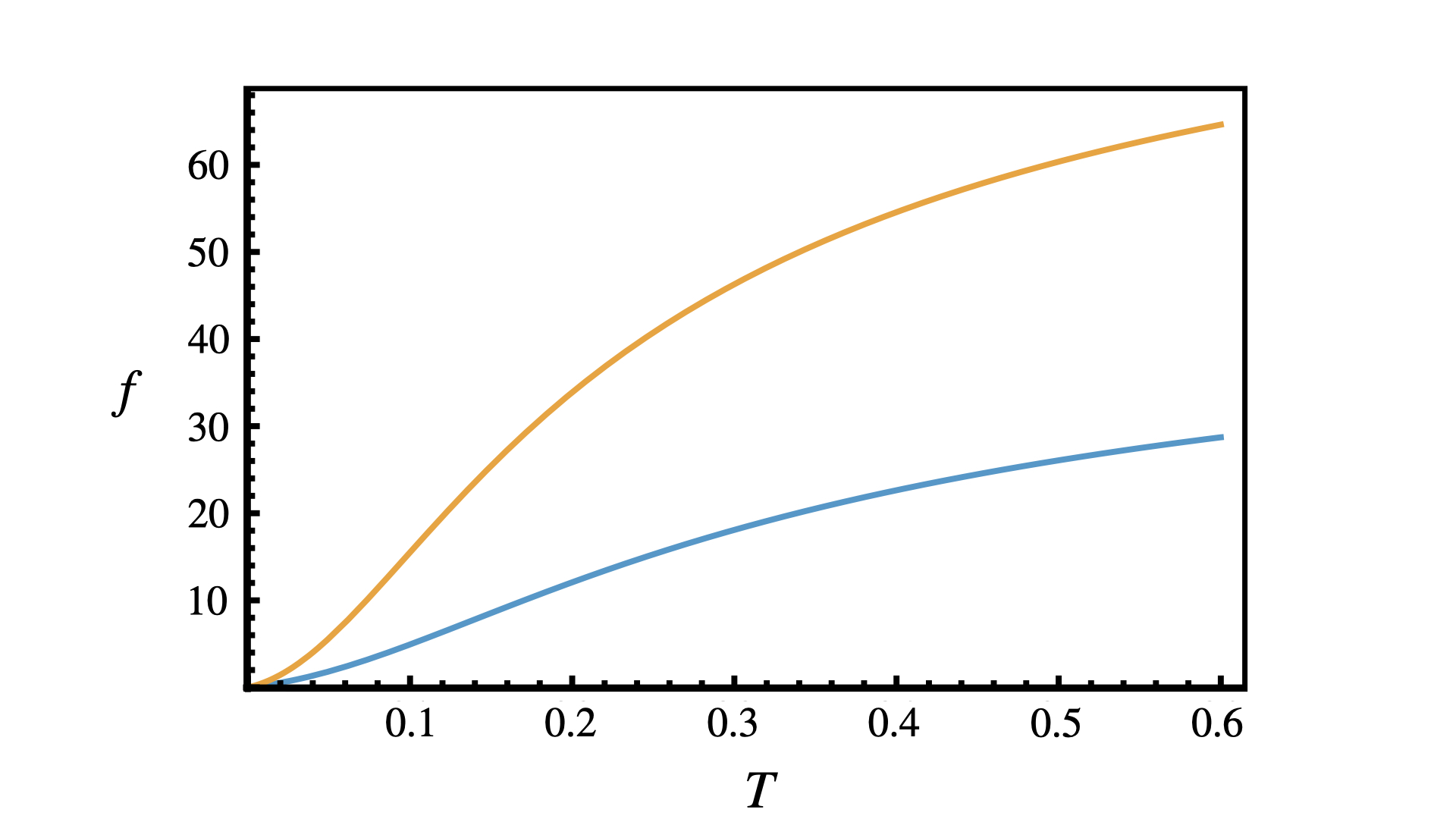}
        \caption{$f(Zm(T)/2T)$ in Eq.~(\ref{sigma}) vs. $T$ at the QCP. The parameters are $t'=0.3$, $t=2K=1$ and $\mu = 0.1$ ($Zm \approx -1.6 +(2.56 +100T^2)^{1/2}$; orange) and $\mu =0.2$ ($Zm \approx -1.5 +(2.25 + 196 T^2)^{1/2}$; blue).}
        \label{fig:Res}
\end{figure}

Since we are mostly interested in the behavior at the quantum critical point, we fix the parameters such that $m(T=0) =0$. Numerical calculations show that 
the temperature dependence of $m = 1/J - \Pi(0,0;T)$ interpolates between quadratic at smallest $T$ and linear at larger $T$. For example, for $t'= 0.3$, $t=2K=1$ we have for $\mu =0.1$ $m \approx -1.6 +(2.56 +100T^2)^{1/2}$ and for $\mu =0.2$ $m \approx -1.5 +(2.25 + 196 T^2)^{1/2}$ 
The corresponding $T$-dependence of the resistivity is depicted in Fig.~\ref{fig:Res}.

In addition to $\sigma_{\mathrm{AL}}$, there is a quasiparticle contribution to the conductivity (``DOS'' contribution); the phase coherence ``Maki-Thompson'' contribution in the present case is absent. At criticality, the quasiparticle lifetime is controlled by scattering from the overdamped hybridization mode; close to the AF Brillouin-zone boundary (small $k_{\perp}$) quasiparticles are strongly overdamped, while further from the boundary the damping crosses over to an approximately linear frequency dependence. We leave a careful study of the transport effects, including the question about the actual source of momentum relaxation processes (Umklapp-scattering) to future investigations.

\section{Conclusions}

We have presented a square-lattice model for a stable spin liquid with a Majorana Fermi surface, coupled to a Fermi sea. Our results provide a concrete example of exchange coupling between conduction electrons and an underlying spin liquid  transforming the electron Fermi-surface volume without symmetry breaking. While the original model, Eq.~\eqref{eq:MainModel} resembles 
Sachdev's ancilla model\textit{et al.}~\cite{ALM1} for the pseudogap phase of the cuprates there are certain key differences.

First, in the large Fermi surface phase, our model  hosts a gapless $\mathbb Z_2$ quantum spin-liquid with a spinon Fermi surface which contributes to the thermodynamics. { In the limit of static gauge fields, both phases of our model appear to be fractionalized. In the decoupled phase with the large Fermi surface we have electrically neutral fermions (spinons) with spin 1/2, in the small Fermi surface state the spinons hybridize with the electrons. However, thermal and/or quantum fluctuations of gauge and hybridization fields will destroy the long-range order at longest spatiotemporal scales.} %, but fractionalized excitations change their statistics and become collective bosonic excitations with charge $e$ and spin 1/2. At finite temperatures they do not condense due to the non-Abelian thermal fluctuations. } %is also fractionalized. 
Second, this model is analytically tractable using conventional many-body techniques, as the $f$ fermions are unconstrained in Eq.~\eqref{HF}.  

A critical point separates the two regimes with different conduction-electron Fermi-surface volumes. In the hybridized regime, the electrons and spinons form a single common Fermi sea. If the hybridization is lost, the spinons form a square Fermi surface (as in the $J\!=\!0$ limit), giving a large density of states and an enhanced Sommerfeld coefficient. However, for any finite $J$ the induced RKKY interaction between spinons is expected to cut off this singular behavior, and may drive magnetic order and open a gap.

Since our results reproduce certain salient features of cuprate phenomenology, it is tempting, following the logic of Refs.~\cite{ALM1,ALM2}, to consider them as  universal properties of the FL$^*$. Most notably, the normal-state Green's function takes the YRZ form.  The FL$^*$ state which we identify with the pseudogap develops strong diamagnetic fluctuations over a broad temperature range. There is also a quantum critical point between the FL and FL$^*$ phases with a logarithmically divergent Sommerfeld coefficient. By contrast, our results for the resistivity and the quasiparticle lifetime at criticality are less encouraging: although there is an extended regime of $R(T)\propto T$, noticeable deviations appear both at low and at high temperatures. This may be an artifact of the RPA treatment.

A central question is whether these results have more than an accidental connection to the cuprates. In this regard, non-perturbative studies of the Hubbard model find a large electron self-energy near the antiferromagnetic Brillouin-zone boundary, consistent with the YRZ picture\cite{Hidden,Imada, MonteCarlo}. In Ref.~\cite{Hidden}, cluster DMFT calculations of the single-particle Green's function were fit by a form similar to Eq.~(\ref{HF}). Refs.~\cite{Hidden,Imada} interpret these results as evidence for a ``hidden fermion'', namely our $f$ particle.

\section*{Acknowledgments}
AT is grateful to G. Kotliar, A. Weichselbaum and W. Yin for valuable discussions. AP would like to thank Andreas Gleis for valuable discussions.  This work was supported by Office of Basic Energy Sciences, Materials
Sciences and Engineering Division, U.S. Department of Energy (DOE)
under Contracts No. DE-SC0012704 (AMT) and DE-FG02-99ER45790 (PC). AP was supported by the U.S. Department of Energy, Office of Science, Basic Energy Sciences, Materials Sciences and Engineering Division. Support for this research was provided by the Office of the Vice Chancellor
for Research and Graduate Education at the University
of Wisconsin–Madison with funding from the Wisconsin
Alumni Research Foundation (EJK). {This work was performed in part at the Aspen Center for Physics, which is supported by National Science Foundation grant PHY-2210452. All authors acknowledge the hospitality of the Aspen Center for Physics.}

\appendix

\section{Yao-Lee-Kondo model.}

\label{app:YLDetails}

The starting point for our considerations, see Eq.~\eqref{eq:MainModel} is the generalized Yao-Lee model introduced in \cite{Vojta2}, which we here discuss in more details. The model is defined on a lattice with coordination number four (this includes square lattice) and is formulated in terms of gamma matrices $\lambda^\gamma$ and their commutators $\lambda^{\alpha\beta} = \frac{i}{2}[\lambda^{\alpha},\lambda^{\beta}]$
 \bea
&& H_{\rm YL} = \sum_{<i,j>}\sum_{\gamma=1}^4K^{ij}
_{\gamma}\Big[\lambda_i^{\gamma}\lambda_j^{\gamma} + \sum_{\beta =5}^7\lambda^{\gamma\beta}_i\lambda_j^{\gamma\beta}\Big] \nonumber\\
&&  - h \sum_{\square} B_\square, \label{Coord4} 
\eea
Note the directional dependent couplings, here encoded in $K_\gamma^{ij}$ as discussed in Sec.~\ref{sec:model}. Model (\ref{Coord4}) can be also be explicitly represented as products of the Pauli matrices: 
\begin{eqnarray}
    (\lambda^1,\lambda^2,\lambda^3) &=&  (\tau^1,\tau^2,\tau^3)\otimes\kappa^y\otimes 1 \cr
    \lambda^4 &=& 1\otimes      \kappa^x\otimes 1\cr
    (\lambda^5,\lambda^6,\lambda^7) &=& 1 \otimes
    \kappa^z \otimes  (\rho^1,\rho^2,\rho^3)
    \end{eqnarray}
 Following the spirit of the ancilla model \cite{ALM1} we identify the top layer of spins with the  $\lambda^{5-7}$, and to delineate these variables, we re-label them as
 \begin{eqnarray}
     (\lambda^5,\lambda^6,\lambda^7) = (\Gamma^x,\Gamma^y,\Gamma^z)\equiv \bm{\Gamma}
 \end{eqnarray}
 This variable plays the role of a conventional spin, with one exception - it anticommutes with the ancillary Cliffords $\lambda^{1,2,3,4}$. We can then rewrite the spin liquid in the following form
 \begin{eqnarray}
H_{YL}=\sum_{<i,j>}\sum_{\gamma=1}^4K^{ij}_{\gamma}\lambda_i^{\gamma}\lambda_j^{\gamma}
     \Big[1- \bm{\Gamma}_i\cdot \bm{\Gamma}_j\Big] - h \sum_{\square} B_\square. 
 \end{eqnarray}
Finally, we choose the bond-variables $K^{ij}_{\gamma} = -\frac{K}{2} \delta_{\gamma,\gamma_{ij}}$, where $\gamma_{ij}\in(1,2,3,4)$ defines which of the four different bond types the bond belongs to, we obtain 
\begin{equation}
    H_{\rm YL} = -\frac{K}{2}\sum_{\langle i,j\rangle}(\lambda^{\gamma_{ij}}_i\lambda^{\gamma_{ij}}_j)
\Bigl[1- \bm{\Gamma}_i\cdot \bm{\Gamma}_j\Bigr] - h\sum_{\square} B_{\square},
\end{equation}
as in Eq.~\eqref{eq:YLKappaTauRho}.

The model Eq.~\eqref{Coord4} can be fermionized as follows, 
\bea
\lambda^{\alpha} = \ri b^8b^{\alpha}, ~~\lambda^{\alpha\beta} = \ri b^{\alpha}b^{\beta},\label{rules}
\eea
where the Majorana fermions $b^{\alpha}$ satisfy  the Clifford algebra
$\{ b^{\alpha},b^{\beta}\} = 2\delta^{\alpha\beta}$. The condition  $\prod_{\alpha=1}^{7}\lambda^{\alpha} = i$ then enforces the constraint
\bea
b_1b_2...b_8 = -1.
\eea
The resulting Hamiltonian is the $\mathbb Z_2$ gauge theory with static gauge fields: 
\begin{align}
H_{\rm YL} &= { \frac{\ri}{2}} K\sum_{<i,j>}\sum_{\alpha=5}^8 \hat u_{ij}b_i^{\alpha}b_j^{\alpha} -h \sum_{\square} B_\square, \notag\\
& \equiv {\frac{\ri}{2} K \sum_{i} \sum_{\alpha=5}^8 \sum_{\mu = x,y} \hat u_{i,i + \hat e_\mu}b_i^{\alpha}b_{i + \hat e_\mu}^{\alpha} } \notag \\
&-h \sum_{\square} B_\square \label{YL3} 
\end{align}
where  

\bea
\hat u_{ij} = \ri b^{\gamma_{ij}}_ib^{\gamma_{ij}}_j, ~~\gamma_{ij}\in (1,2,3,4) \label{v}, 
\eea
is the $\mathbb Z_2$ gauge field and $\hat e_{x,y}$ are lattice vectors in $x, y$ direction, respectively.  Hence the Yao-Lee model (\ref{YL3}) is a model with four species of noninteracting Majorana fermions in the background of a static $\mathbb Z_2$ gauge field. {We also remark that the flux $B_{\square} = \prod_{(ij)\in\partial\square} \hat u_{ij}$ can be explicitly presented in terms of the $\mathbb Z_2$ gauge fields.} 

  We define the Dirac fermions and we explicitly use the bipartiteness of the square lattice:  if $j$ is on the $A$ sublattice
\begin{subequations}
\bea
f_{\uparrow,j} = \ri\Big(\frac{b^5 +\ri b^6}{2}\Big), ~~f_{\downarrow,j} = \ri \Big(\frac{b^7 +\ri b^8}{2}\Big).\label{Dirac}
\eea
 In contrast, if $j$  is a site on the $B$ sublattice 
\bea
f_{\uparrow,j} = \Big(\frac{b^5 +\ri b^6}{2}\Big), ~~f_{\downarrow,j} =  \Big(\frac{b^7 +\ri b^8}{2}\Big).\label{Dirac}
\eea
\end{subequations}

{Using this expression we find
\begin{align}
    &\sum_{\sigma, \pm} \sum_{\mu = x,y} \sum_{i \in A} \hat u_{i, i \pm \hat e_\mu} (f^\dagger_{i \sigma} f_{i \pm \hat e_\mu, \sigma} + H.c.) 
    \notag \\
    & = \frac{- i}{2} \sum_{\mu = x,y} \sum_{i \in A} \hat u_{i, i \pm \hat e_{\mu} } \sum_{\alpha = 5}^8 b^\alpha_i b^\alpha_{i \pm \hat e_\mu} \notag \\
    & = - \frac{i}{2} \sum_{\mu = x,y} \sum_i \sum_{\alpha = 5}^8 \hat u_{i, i + \hat e_\mu} b^\alpha_i b^\alpha_{i + \hat e_\mu}.
\end{align}}

We can remove the requirement that $i\in A$ is on the $A$ sublattice by introducing the index ordering notation 
\begin{equation}
   u_{(i,j) }= \left\{ 
   \begin{array}{rl} 
   u_{ij} & i\in A, j\in B\\
   u_{ji} & j\in A, i \in A
    \end{array} \right.
\end{equation}
Whereas $u_{ij}=-u_{ji}$ is an odd function of indices,  $u_{(i,j)}= u_{(j,i)}$ is an even function of indices. 
 We now couple the electrons to 
 the spin-like degree of freedom of the spin liquid 
 defined in terms of the ${(1 + \kappa_z) \bm{\Gamma}}$. 
 We find that the resulting spin density in the YL sector is given by  \bea f\dg{\bm{\sigma}}f = \frac{1}{2}\Big((1 +\kappa_z) \bm{\Gamma}^i\Bigr),\eea where $\kappa_z=-i \Gamma^5\Gamma^6\Gamma^7$. In the main text, we simply use the notation $\kappa = \kappa_z$.

  Substituting the previous results into (\ref{YL3}) we obtain Hamiltonian (\ref{HF}) with $\hat u_{ij} = \hat u_{ji}$:
 \bea 
H &=& -\sum_{i,j} t_{ij}\bigl(c^\dagger_{i,\sigma} c_{j,\sigma} + \mathrm{H.c.}\bigr)
+ J\sum_i (c^\dagger \boldsymbol{\sigma} c)_i \cdot  (f^\dagger \boldsymbol{\sigma} f)_i\nonumber\\ 
&& {-} K\sum_{\langle i,j\rangle} \hat u_{(i,j)}
\bigl(f^\dagger_{i,\sigma} f_{j,\sigma} + \mathrm{H.c.}\bigr)
- h\sum_{\square} B_{\square}.
\label{HF2}
\eea

This concludes the derivation of~\eqref{HF} from Eq.~\eqref{eq:MainModel}, in particular we derived the explicit form of the Kondo coupling.

\section{Derivation of Ginzburg Landau free energy}
\label{app:GLDetails}

\subsection{Foundations}

For the fixed gauge $u_{(i,j)} = 1$, we obtain the following  mean field Hamiltonian
\begin{align}
    \hat H_{\rm MF} &= \sum_{k} [c^\dagger_{k\sigma} \epsilon_{k} c_{k, \sigma} + f^\dagger_{k\sigma} K_{k} f_{k, \sigma}] \notag \\
    &+ \sum_{k} [V c^\dagger_{k\sigma} f_{k+Q, \sigma} + H.c.] \notag \\
    & +\sum_k [\Delta c^\dagger_{k,\sigma} (i \sigma_y)_{\sigma \sigma'} f^+_{-k, \sigma'} +H.c. ]
\end{align}
We introduce the spinor 
\begin{equation}
    \Psi_k = \left ( \begin{array}{c}
         c_{k, \uparrow} \\
         c\dg_{-(k+Q), \downarrow} \\
         f_{k+Q, \uparrow} \\
         f\dg_{-k, \downarrow}
    \end{array}\right)
\end{equation}
to rewrite
\begin{equation}
    \hat H = \sum_k \Psi_k^\dagger \left (\begin{array}{cccc}
        \epsilon_k & 0 & V & \Delta \\
        0 & - \epsilon_{- (k + Q)} & \Delta^* & - V^* \\
        V^* & \Delta & K_{k + Q} & 0 \\
        \Delta^* & -V & 0 & - K_{-k}
    \end{array} \right ) \Psi_k,
\end{equation}
where $Q=(\pi,\pi)$ { and following \eqref{disp},
\bea
\label{disp2}
&& \epsilon_{k} = - 2t(\cos k_x +\cos k_y) +4t'\cos k_x\cos k_y-\mu, \label{t}\cr
&& K_{k} = -{2}K(\cos k_x +\cos k_y). 
\eea }

Observing that the off-diagonal blocks are proportional to an SU(2) matrix { $U$}
\begin{equation}
   \left ( \begin{array}{cc}
        V & \Delta \\
        \Delta^* & - V^* 
    \end{array} \right ) = i\sqrt{\vert V \vert^2 + \vert \Delta \vert^2} U,
\end{equation}
 { 
and noting that he spinon dispersion is particle-hole symmetric, $K_{-(k +Q)} = -K_k$, we can compactly write  
\begin{equation}
     \hat H = \sum_k \Psi_k^\dagger\begin{pmatrix}
        \epsilon_k \underline{1}_2 + \delta_k \tau_3 & i\Phi U\cr
        -i \Phi U\dg & -K_{k}1_2
    \end{pmatrix}  \Psi_k,
\end{equation}
where $\Phi = \sqrt{\vert V \vert^2 + \vert \Delta \vert^2}$.
Here  
$\tau_3$ is the third Pauli matrix and we have divided the electron dispersion into particle-hole symmetric and antisymmetric components 
$\epsilon_k = \epsilon^{(0)}_k + \delta_k$, where 
\begin{eqnarray}
\epsilon^{(0)}_k &=& - 2t [\cos(k_x) + \cos(k_y)],\cr  \delta_k &=& 4t' \cos(k_x) \cos(k_y) - \mu,\end{eqnarray}
such that $\epsilon^{(0)}_k = - \epsilon^{(0)}_{k+Q}$ and $\delta_k = \delta_{k+Q}$.
We can now  absorb the matrix $U$ into a Bogoliubov transformation
\begin{equation}
    \Psi_{k}\rightarrow \begin{pmatrix} 1_2 & 0 \cr
    0 & -i U\dg 
    \end{pmatrix}\Psi_k,\  \Psi\dg_{k}\rightarrow \Psi\dg_k\begin{pmatrix} 1_2 & 0 \cr
    0 & i U
    \end{pmatrix},
\end{equation}so that Hamiltonian is block diagonal 
\begin{equation}
 \hat H = \sum_k \Psi_k^\dagger\begin{pmatrix}
        \epsilon^{(0)}_k \underline{1}_2 + \delta_k \tau_3 & \Phi\underline{1}_2\cr
        \Phi \underline{1}_2 & -K_{k}1_2
    \end{pmatrix}  \Psi_k.
\end{equation}
The mean-field bands are thus
\begin{equation}
    E_{\xi,\tau}(k) =  E^{(-)}_{\tau}(k)+ \xi\sqrt{ E^{(+)}_{ \tau}(k)^2 + \Phi^2} \label{B9}
\end{equation}
where $\xi=\pm 1$, $\tau = \pm 1$ and
$
     E^{(\pm)}_{\tau}(k) = \frac{1}{2}(\epsilon_{k}^{(0)} + \tau \delta_{k} \pm  K_k)
$.
%In particular this implies that the effective Ginzburg-Landau functional is perfectly SU(2) symmetric. 

\begin{figure}
        \centering
    \includegraphics[width=0.6\linewidth]{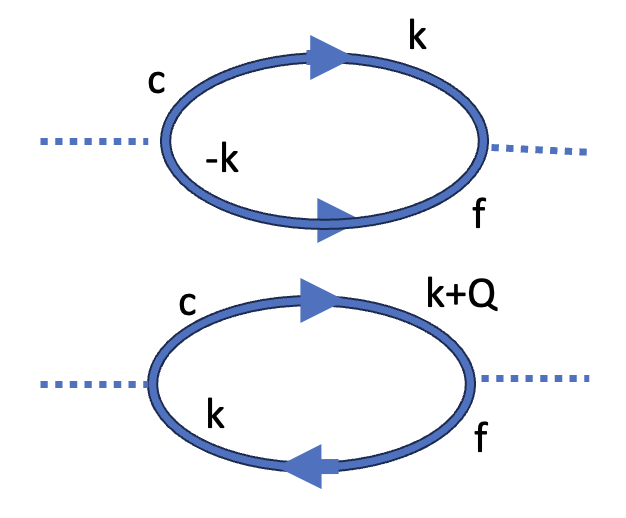}
    \caption{{Bubble diagrams for the $V-V$ and $\Delta-\Delta$ susceptibilities. Due to the particle-hole symmetry of the $f$-particles these diagrams are equal.} %along the dotted line trajectory shown in {\bf a.}
}
    \label{fig:bubbles}
\end{figure}
\begin{widetext}
 {   
The free energy density contribution of the fermions is  then
\begin{align}
    f & = - k_BT \sum_{\xi, \tau} \int_{k} \ln (1 + e^{ - \beta E_{\xi, \tau}(k)}) + \frac{\Phi^2}{J} \cr
    & = \int_k \left\{\sum_\tau (\epsilon_k^0+ \tau \delta_k) - k_B T \sum_{\xi,\tau}  \ln\biggl[2\cosh\left(\frac{\beta E_{\xi,\tau}(k)}{2}\right)\biggr]\right\}+ \frac{\Phi^2}{J} \cr
    &=- k_B T \sum_{\xi,\tau} \int_k \ln\biggl[2\cosh\left(\frac{\beta E_{\xi,\tau}(k)}{2}\right)\biggr]+ \frac{\Phi^2}{J},
\end{align}
where we denote $\int_{k} \equiv \int \frac{d^2k}{(2\pi)^2}$ and have identified 
$\int_k \epsilon_k^{(0)}=0$. 
\begin{comment}Using the identity $2\cosh [(A+B)/2]\cosh [(A-B)/2]= \cosh A  + \cosh B$, where $A = \beta\tilde E_{k,\tau}$ and $B = \beta\sqrt{\Delta \tilde E_{k,\tau}^2 + \Phi^2}$, we then obtain 
\begin{align}
    f & =- k_B T \sum_{\xi,\tau} \int_k \ln\biggl[2\cosh\bigl(\frac{\beta E_{\xi,\tau}(k)}{2}\bigr)\biggr]+ \frac{\Phi^2}{J}\\
    &= - k_B T \sum_{\tau} \int_k \ln\biggl[2(\cosh\bigl(\beta \tilde E_{k,\tau}\bigr) + \cosh\bigl(\beta \sqrt{\Delta \tilde E_{k,\tau}^2 + \Phi^2}\bigr)\biggr] + \frac{\Phi^2}{J}\nonumber
\end{align}
\end{comment}
}

\subsection{Quadratic terms in free energy}

{ The logarithmically divergent diagrams are  given on Fig. \ref{fig:bubbles}}.

The quadratic terms in $\Phi$ in the free energy are
\begin{align}
    f^{(2)}  = - \sum_\tau \int_k \frac{\Phi^2}{2 (\epsilon_k + \tau \delta_k + K_k)}
    \times  \left[\tanh\left(\frac{\epsilon_k + \tau \delta_k}{2T}\right)+\tanh\left(\frac{K_k}{2T}\right)\right]+ \frac{\Phi^2}{J}, 
\end{align}
This expression is  most easily evaluated in the case when $t' = 0$ where we can use the dimensionless density of states
\begin{equation}
    \nu(x) = \int_k \delta [x + 2(\cos k_x + \cos k_y)]=\frac{2}{\pi^2}K\left({1-\left(\frac{x}{4}\right)^2}\right) \stackrel{\vert x \vert \ll 1}{\simeq} \frac{2}{\pi^2}\ln\left(\frac{16}{\vert x\vert}\right),
\end{equation}
 %Alexei please double check the fine detail of these calculations }
 where $K(x)$ is the elliptic integral of the first kind,
to express
\begin{align}\label{ftT}
    f^{(2)} =  - \Phi^2\sum_\tau \int_0^4 dx \frac{ \nu(x) }{ (t+K)x + \tau \mu}\times  \left[\tanh(\frac{t x + \tau\mu}{2T})+\tanh(\frac{K x}{2T})\right]+ \frac{\Phi^2}{J}.
\end{align}
%{\bf AT: }
%{\bf \color{red} PC thinks pre-factor is 2, not 4, because we get one factor of 2 from combining postive and negative $x$, and another factor of two from combining the two tanh or sgn functions, this is four, but we must divide by the factor of 2 in the denominator, hence 2 x 2/2 = 2 as the prefactor. } 
At zero temperature and finite $\mu$ this is 
\begin{equation}
    f^{(2)} =  - 2
 {\Phi^2}\left\{\int_0^{4}dx\frac{\nu(x)}{x(t+K) + \mu}+\int_{\mu/t}^{4} dx \frac{ \nu(x) }{x{(t+K)}- \mu }  \right\}+ \frac{\Phi^2}{J}.
\end{equation}
% \bf 
%PC Please note that the variable x is dimensionless.  The upper cutoff is 4, not $\Lambda$. %In excruciating detail, m
Making the change of variable $u= \ln (16/x)$, we obtain
\begin{equation}
    f^{(2)} = -\frac{4\Phi^2}{\pi^2} \left\{ \int _{\ln 4}^{\infty} du \frac{u}{(t+K) + \frac{\mu}{16}e^u}
    +  \int _{\ln 4}^{\ln \frac{16}{\mu/t}} du  \frac{u}{(t+K) - \frac{\mu}{16}e^u}\right\}+ \frac{\Phi^2}{J}.
\end{equation}
The integrands scale like u for small u, but the exponentials in the denominators become important when $|(t+K)| \sim \mu e^u / 16$, i.e when $u= u_1\sim \ln (16(t+K)/\mu)$. This becomes the upper cutoff in the first integral. However, in the second integral, we observe that the existing cutoff is already smaller than $u_1$, so the second cutoff is unaffected.  We  thus calculate: 
\begin{eqnarray}\label{f2mu}
 &&f^{(2)} \approx  -\frac{4\Phi^2}{\pi^2(t+K)}\left[\int ^{\ln (16(t+K)/\mu)}_{\ln 4} + \int _{\ln 4}^{\ln \frac{16}{\mu/t}}\right] du u + \frac{\Phi^2}{J} \cr
  &&  =   - \frac{2 \Phi^2}{\pi^2(t + K)} \left[\ln^2\left(\frac{16(t+K)}{|\mu|}\right) + \ln^2\left(\frac{16t}{|\mu|}\right)- 2 \ln^2 4 \right]  + \frac{\Phi^2}{J},\cr
  &&  =   - \frac{4 \Phi^2}{\pi^2(t + K)} \left[\ln^2\left(\frac{16\sqrt{(t+K)t}}{|\mu|}\right) + \frac{1}{4}\ln^2\left(\frac{t+K}{t}\right)-  \ln^2 4 \right]  + \frac{\Phi^2}{J},
\end{eqnarray}
The error in this expression is  of order $2 \mu \Phi^2/ (16(t+K)^2) $.
%{\color{green} With logarithmic accuracy we obtain 
%{\bf Revised version that Piers does not agree with.  }
%\begin{eqnarray}
 %  f^{(2)} =   \frac{\Phi^2}{J}- \frac{{ 8} \Phi^2}{\pi^2(t + K)} \Big\{\ln^2\left(\frac{ 16 t}{|\mu|}\right) + \ln\left(\frac{ 16 t}{|\mu|}\right)\ln\left(\frac{(t+K)^2}{Kt}\right) +...\Big\}% \ln^2 \left(\sqrt{\frac{t+K}{t}}\right)- \ln^2 4\right]  ,
%\end{eqnarray}
%}
This leads to an instability, { provided the spin coupling $J>J_c$ where
\begin{equation}
J_c = \frac{\pi^2(t + K)}{4} \Big\{\ln^2\left(\frac{16\sqrt{(t+K)t}}{|\mu|}\right) + \frac{1}{4}\ln^2\left(\frac{t+K}{t}\right)-  \ln^2 4 \Big\}^{-1}
\end{equation}
}
or alternatively, when
\begin{eqnarray}
 \mu < \mu_c = 16\sqrt{t(t+K)}\exp{\Big[-\sqrt{\pi^2(t+K)/4J + \ln^2 4 -\frac{1}{4}\ln^2(1+K/t) }\Big]}. \label{muc}
\end{eqnarray}

To obtain an estimate of the mean-field transition temperature, we assume that $T \gg \mu$. In this case, the upper-cutoff in \eqref{f2mu} is determined by the temperature inside the two tanh functions in \eqref{ftT}, so that in this limit 
\begin{eqnarray}
   f^{(2)}(T) =   - \frac{2 \Phi^2}{\pi^2(t + K)} \left[\ln^2\left(\frac{16 t}{T}\right) + \ln^2\left(\frac{16K}{T}\right)- 2 \ln^2 4 \right]  + \frac{\Phi^2}{J},\cr
\end{eqnarray}
giving rise to the mean-field transition temperature
\begin{eqnarray}\label{mfTc}
T_c = 16\sqrt{tK}\exp{\Big[-\sqrt{\pi^2(t+K)/4J + \ln^2 4 -\frac{1}{4}\ln^2(t/K) }\Big]}. \label{tc}
\end{eqnarray}

\end{widetext}
}
%, where 
%{\color{green}
%\begin{equation}\label{mucr}
%\mu_c = 16\frac{(t+K)^2}{\sqrt{tK}}e^{-\frac{\pi}{2} \sqrt{\frac{t+K}{2J}}}.
%\end{equation}
%}
    %\mu_c = 16(t+K)e^{-\frac{\pi}{2} \sqrt{\frac{t+K}{J}+ \ln^2 4-\ln^2 \left(\sqrt{\frac{t+K}{t}}\right)}}.
%\end{equation}
%{\bf AT version:}
%\begin{equation}
 %   \mu_c = 16(t+K)(t/K)^{1/2}e^{-\frac{1}{2} \sqrt{\pi^2\frac{t+K}{2J}- \ln^2 (K/t)}}.
%\end{equation}

%By contrast, at finite temperature and $\mu=0$ we find
%\color{green}
%\begin{equation}
 %   f^{(2)} = - \frac{8 \Phi^2}{\pi^2(t + K)} \left [\ln^2 \left ( {\frac{8t}{T}}\right ) + \ln\left(\frac{8t}{T}\right) \left (\frac{K}{t}\right ) \right ]+\frac{\Phi^2}{J},
%\end{equation}

%\begin{equation}
%    f^{(2)} = - \frac{8 \Phi^2}{\pi^2(t + K)} \left [\ln^2 \left ( {\frac{2T}{\sqrt{Kt}}}\right ) + \ln^2 \left ( \sqrt{\frac{K}{t}}\right ) \right ]+\frac{\Phi^2}{J},
%\end{equation}
%leading to a {\color{green} mean field critical temperature $T_c \sim \mu_c$} {\color{red}What does this mean?}
%{ mean-field critical temperature }
%\begin{equation}\label{mftc}
 %   T_c = \frac{\sqrt{Kt}}{2} \exp\biggl[{ -\sqrt{\frac{\pi^2 (t + K)}{8J}- \ln^2 \left ( \sqrt{\frac{K}{t}}\right )}}\biggr].
%\end{equation}
\section{Visons in the Principal Chiral Field model}

We will show, as promised, that these fluctuations do not modify the correlation length. For this end we chose a particular gauge for the $u_{ij}$ field setting its value to 1 except for $(na_x, 2ma_y);(n+1)a_x, 2ma_y)$ links where it can be either $+1$ or $-1$. This way  every plaquette can have flux $\pm 1$ and the re is no over-summation. Summing on $u_{ij}$ on these links and expanding in $\exp(-\kappa/T)$ we obtain the following contribution to the partition function:
\bea
&& Z_{(x,2y);(x+a,2y)} = \\
&& \exp\Big[ \frac{\rho}{2T}\mbox{Tr}(U_{x,2y}U\dg_{x+a,2y}+ H.c)\Big] +\nonumber\\
&& \re^{-\kappa/T}\exp\Big[ -\frac{\rho}{2T}\mbox{Tr}(U_{x,2y}U\dg_{x+a,2y}+ H.c.)\Big] \approx \nonumber\\
&& \exp\Big\{ -\frac{\rho a^2}{2T}\mbox{Tr}(\p_x U\p_x U\dg)\Big[1 +2\exp[(-\kappa-2\rho)/T)]\Big]\Big\} .\nonumber
\eea
So the vison contribution in this limit is reduced to a renomalization of stiffness.  In the continuum limit the model becomes the principle chiral field one. 

\bibliography{flstar.bib}

@article{Sereni,
author = {Sereni, J G and Geibel, C and Berisso, M G and Hellmann, P},
title = {{\sl Scaling of the C$_p\sim T 1nT$ dependence in Ce systems}},
journal = {Physica B: Condensed Matter Physics},
pages = {580-582},
volume = {230-232},
year = {1997}
}

@article{flintcoleman2009, 
 title={Symplectic Nand time reversal in frustrated magnetism}, 
 volume={79}, ISSN={1098-0121}, url={http://link.aps.org/doi/10.110
 3/PhysRevB.79.014424}, 
 DOI={10.1103/physrevb.79.014424}, 
 number={1}, 
 journal={Phys. Rev. B}, 
 author={Flint, Rebecca and Coleman, P and Coleman, Piers}, 
 year={2009}, 
 month=jan, 
 pages={014424}}

@article{AbanovSchmalian2003,
  title={Quantum-critical theory of the spin-fermion model and its application to cuprates: Normal state analysis},
  author={Abanov, Artem and Chubukov, Andrey V and Schmalian, J{\"o}rg},
  journal={Advances in Physics},
  volume={52},
  number={3},
  pages={119--218},
  year={2003},
  publisher={Taylor \& Francis}
}

@article{MetlitskiSachdev2010,
  title = {Quantum phase transitions of metals in two spatial dimensions. II. Spin density wave order},
  author = {Metlitski, Max A. and Sachdev, Subir},
  journal = {Phys. Rev. B},
  volume = {82},
  issue = {7},
  pages = {075128},
  numpages = {30},
  year = {2010},
  month = {Aug},
  publisher = {American Physical Society},
  doi = {10.1103/PhysRevB.82.075128},
  url = {https://link.aps.org/doi/10.1103/PhysRevB.82.075128}
}

@article{Millis1993,
  title = {Effect of a nonzero temperature on quantum critical points in itinerant fermion systems},
  author = {Millis, A. J.},
  journal = {Phys. Rev. B},
  volume = {48},
  issue = {10},
  pages = {7183--7196},
  numpages = {0},
  year = {1993},
  month = {Sep},
  publisher = {American Physical Society},
  doi = {10.1103/PhysRevB.48.7183},
  url = {https://link.aps.org/doi/10.1103/PhysRevB.48.7183}
}

@article{AffleckMarston1988,
  author = {Affleck, Ian and Marston, J. Brad},
  title = {Large-N limit of the Heisenberg-Hubbard model: Implications for high-Tc superconductors},
  journal = {Physical Review B},
  volume = {37},
  number = {7},
  pages = {3774--3777},
  year = {1988},
  doi = {10.1103/PhysRevB.37.3774}
}

@article{ReadSachdev1990,
  author = {Read, N. and Sachdev, Subir},
  title = {Spin-Peierls, valence-bond solid, and N\'eel ground states of low-dimensional quantum antiferromagnets},
  journal = {Physical Review B},
  volume = {42},
  number = {7},
  pages = {4568--4589},
  year = {1990},
  doi = {10.1103/PhysRevB.42.4568}
}

@article{TrebstHickey2022,
  author = {Trebst, Simon and Hickey, Ciar\'an},
  title = {Kitaev Materials},
  journal = {Physics Reports},
  volume = {950},
  pages = {1--37},
  year = {2022},
  doi = {10.1016/j.physrep.2021.11.003}
}

@article{CPT1, title={Solvable 3D Kondo Lattice Exhibiting Pair Density Wave, Odd-Frequency Pairing, and Order Fractionalization}, volume={129}, ISSN={0031-9007}, DOI={10.1103/physrevlett.129.177601}, abstractNote={The Kondo lattice model plays a key role in our understanding of quantum materials, but a lack of small parameters has posed a long-standing problem. We present a three-dimensional S=1/2 Kondo lattice model describing a spin liquid within an electron sea. Strong correlations in the spin liquid are treated exactly, enabling a controlled analytical approach. Like a Peierls or BCS phase, a logarithmically divergent susceptibility leads to an instability into a new phase at arbitrarily small Kondo coupling. Our solution captures a plethora of emergent phenomena, including odd-frequency pairing, pair density wave formation and order fractionalization. The ground-state state is a pair density wave with a fractionalized charge e, S=1/2 order parameter, formed between electrons and Majorana fermions.}, number={17}, journal={Physical Review Letters}, author={Coleman, Piers and Panigrahi, Aaditya and Tsvelik, Alexei}, year={2022}, pages={177601} }

@article{CPT2, title={Breakdown of order fractionalization in the CPT model}, volume={110}, ISSN={2469-9950}, DOI={10.1103/physrevb.110.104520}, abstractNote={We present an analysis of the half-filled CPT model, an analytically tractable Kondo lattice model with Yao-Lee spin-spin interactions on a 3D hyperoctagon lattice, proposed by Coleman, Panigrahi, and Tsvelik. Previous studies have established that the CPT model exhibits odd-frequency triplet superconductivity and order fractionalization. Through asymptotic analyses in the small-J and large-J Kondo coupling limits, we identify a quantum critical point at Jc, marking a transition from a superconductor to a Kondo insulator. By estimating the vison gap energy to account for thermal gauge fluctuations, we determine the energy scales governing the thermal breakdown of order fractionalization. Moreover, at large J the Kondo insulator undergoes orbital decoupling, leading to the formation of a decoupled Kitaev orbital liquid. These findings and analogies with the Z2-gauged XY model lead us to propose a tentative phase diagram for the CPT model at half-filling.}, number={10}, journal={Physical Review B}, author={Panigrahi, Aaditya and Tsvelik, Alexei and Coleman, Piers}, year={2024}, pages={104520} }

@article{senthil03,
  author = {T. Senthil and S. Sachdev and M. Vojta},
  title = {Fractionalized Fermi Liquids},
  journal = {Physical Review Letters},
  volume = {90},
  pages = {216403},
  year = {2003},
  doi = {10.1103/PhysRevLett.90.216403}
}

@article{CPT3, title={Microscopic Theory of Pair Density Waves in Spin-Orbit Coupled Kondo Lattice.}, volume={135}, ISSN={0031-9007}, DOI={10.1103/pdqz-zb8k}, abstractNote={We demonstrate that the discommensuration between the Fermi surfaces of a conduction sea and an underlying spin liquid provides a natural mechanism for the spontaneous formation of pair density waves. Using a recent formulation of the Kondo lattice model that incorporates a Yao Lee spin liquid proposed by the authors, we demonstrate that doping away from half filling induces finite-momentum electron-Majorana pair condensation, resulting in amplitude-modulated pair density waves (PDWs). Our approach provides a precise, analytically tractable pathway for understanding the spontaneous formation of PDWs in higher dimensions and offers a natural mechanism for PDW formation in the absence of Zeeman splitting.}, number={4}, journal={Physical review letters}, author={Panigrahi, Aaditya and Tsvelik, Alexei and Coleman, Piers}, year={2025}, pages={046504} }

@article{sachdev25,
  author = {P. M. Bonetti and M. Christos and A. Nikolaenko and A. A. Patel and S. Sachdev},
  title = {Fractionalized Critical FErmi liquids and the cuprate phase diagram},
  journal = {Reports on Progress in Physics},
  volume ={89},
  pages ={044501},
  year = {2026},
  doi={10.1088/1361-6633/ae530d}
}

@article{bascones2008,
  author = {Bascones, E. and Valenzuela, B.},
  title = {Yang-Rice-Zhang description of checkerboard pattern and autocorrelation of photoemission data
in high-temperature superconductors},
  journal = {Phys. Rev. B},
  volume ={77},
  pages ={024527},
  year = {2008},
  doi={10.1103/PhysRevB.77.024527}
}

@article{Yamaji2025,
  author = {Yicheng Zhong, Fu-Chun Zhang, Kun Jiang},
  title = {Yamaji effect and quantum oscillation in Yang-Rice-Zhang model of underdoped cuprates},
  journal = {arXiv preprint arXiv:2512.10475},
  year = {2025},
  doi={10.48550/arXiv.2512.10475}
}

@article{deHaas1,
  author = { Doiron-Leyraud, N and Proust, C. and LeBoeuf, D. and  Levallois, J. and 
Bonnemaison, J.-B. and  Liang, R. and Bonn, D. and  Hardy, W. and
Taillefer, L. },
  title = {Quantum oscillations and the Fermi surface in an underdoped high-Tc superconductor},
  journal = {Nature},
  volume={447},
  pages={565},
  year = {2007},
  doi={10.1038/nature05872}
}

@article{deHaas2,
  author = {Bariˇsi´c, N. and  Badoux, S. and Chan, M. K. and  Dorow, C. and Tabis, W. and 
Vignolle, B. and Yu, G. and B´eard, J. and Zhao, X. and Proust, C. {\it et al.}},
  title = {Universal quantum oscillations in the underdoped cuprate superconductors},
  journal = {Nature Physics},
  volume={9},
  pages={761},
  year = {2013},
  doi={10.1038/nphys2792}
}

@article{Wiegmann1984,
  title={On the theory of nonabelian Goldstone bosons in two dimensions; exact solution of the SU(N) $\otimes$ SU(N) nonlinear sigma model},
  author={P. B. Wiegmann},
  journal={Phys. Lett. B},
  volume ={141},
  pages={217},
  year={1984},
  doi ={10.1016/0370-2693(84)90205-3}
}

@article{harrison25,
  author = {M. K. Chan and K. A. Schreiber and O. E. Ayala-Valenzuela and E. D. Bauer and A. Shekhter and N. Harrison},
  title = {Observation of the Yamaji effect in a cuprate superconductor},
  journal = {Nature Physics},
volume ={21},
pages ={1753},
  year = {2025},
doi ={10.1038/s41567-025-03032-2}
}

@article{senthil,
  author = {T. Senthil and S. Sachdev and M. Vojta},
  title = {Fractionalized Fermi Liquids},
  journal = {Physical Review Letters},
  volume = {90},
  pages = {216403},
  year = {2003},
doi ={10.1103/PhysRevLett.90.216403},
  eprint = {cond-mat/0209144}
}

@article{senthil2,
  author = {T. Senthil and M. Vojta and S. Sachdev},
  title = {Weak magnetism and non-Fermi liquids near heavy-fermion critical points},
  journal = {Physical Review B},
  volume = {69},
  pages = {035111},
  year = {2004},
doi ={10.1103/PhysRevB.69,035111},
  eprint = {cond-mat/0305193}
}

@article{Ashvin,
  author = {A. Paramekanti and A. Vishwanath},
  title = {Extending Luttinger’s theorem to Z2 fractionalized phases of matter},
  journal = {Physical Review B},
  volume = {70},
  pages = {245118},
  year = {2004},
doi ={10.1103/PhysRevB.70.245118},
  eprint = {cond-mat/0406619}
}

@article{Grover,
  author = {T. Grover and T. Senthil},
  title = {Quantum phase transition from an antiferromagnet to a spin liquid in a metal},
  journal = {Physical Review B},
  volume = {81},
  pages = {205102},
  year = {2010},
doi ={10.1103/PhysRevB.81.205102},
  eprint = {0910.1277}
}

@article{Bonderson,
  author = {P. Bonderson and M. Cheng and K. Patel and E. Plamadeala},
  title = {Topological Enrichment of Luttinger’s Theorem},
  journal = {arXiv e-prints},
  year = {2016},
  eprint = {1601.07902}
}

@article{Tsvelik2016,
  author = {A. M. Tsvelik},
  title = {Fractionalized Fermi liquid in a Kondo-Heisenberg model},
  journal = {Physical Review B},
  volume = {94},
  pages = {165114},
  year = {2016},
doi ={10.1103/PhysRevB.94.165114},
  eprint = {1604.06417}
}

@article{ALM1,
  author = {Y.-H. Zhang and S. Sachdev},
  title = {From the pseudogap metal to the Fermi liquid using ancilla qubits},
  journal = {Physical Review Research},
  volume = {2},
  pages = {023172},
  year = {2020},
doi={10.1103/PhysRevResearch.2.023172},
  eprint = {2001.09159}
}

@article{ALM2,
  author = {Y.-H. Zhang and S. Sachdev},
  title ={Deconfined criticality and ghost Fermi surfaces at the onset of
antiferromagnetism in a metal}, 
journal ={Phys. Rev. B},
volume ={102},
pages ={155124},
year = {2020},
eprint ={2006.01140},
doi={10.1103/PhysRevB.102.155124}
}

@article{kitaev,
author = {Kitaev, Alexei},
title= {Anyons in an exactly solved model and beyond},
journal ={Annals of Physics},
volume ={321},
pages ={2 –111},
year ={2006}
}

@article{YL,
    author ={Hong Yao and Dung-Hai Lee},
title ={Fermionic magnons, non-abelian
spinons, and the spin quantum Hall effect from an exactly solvable spin-1/2 Kitaev model with su(2) symmetry},
journal ={Phys. Rev.Lett},
volume ={107},
pages ={087205},
year ={2011},
doi={10.1103/PhysRevLett.107.087205}
}

@article{Vojta2,
  author = { Sreejith Chulliparambil and Urban F. P. Seifert and  Matthias Vojta and Lukas Janssen and Hong-Hao Tu},
  title = {Microscopic models for Kitaev’s sixteenfold way of anyon theories},
  journal = {Physical Review B},
  volume = {102},
  pages = {201111},
  year = {2020},
doi ={10.1103/PhysRevB.102.201111}
}

@article{Peter,
  author = { M. Khodas and  H.-B. Yang and  J. Rameau and  P. D. Johnson and A.M. Tsvelik},
  title = {Analysis of the Quasiparticle Spectral Function in the Underdoped Cuprates},
  journal = {arXiv},
  volume = {1007.4837},
  pages = {},
  year = {2010},
doi = {10.48550/arXiv.1007.4837}
}

@article{YRZ,
  author = { K.-Y. Yang and  T. M. Rice and F.-Ch. Zhang},
  title = {Phenomenological theory of the pseudogap state},
  journal = {Phys. Rev. B},
  volume = {73},
  pages = {174501},
  year = {2006},
doi = {10.1103/PhysRevB.73.174501}
}

@article{zommerfeld,
  author = { B. Michon and C. Girod and S.Badoux and J. Kačmarčík and Q. Ma  and M. Dragomir and  H. A. Dabkowska and B. D. Gaulin and  J.-S. Zhou and  S. Pyon and  T. Takayama and  H. Takagi and  S. Verret and  N. Doiron-Leyraud and  C. Marcenat and  L. Taillefer and  T. Klein},
  title = {Thermodynamic signatures of quantum criticality in cuprates},
  journal = {Nature},
  volume = {567},
  pages = {218},
  year = {2018},
doi = {10.1038/s41586-019-0932-x}
}

@article{MonteCarlo,
  author = { R. Rossi and F. Simkovic IV and M. Ferrero and A. Georges and A. M. Tsvelik and N. V. Prokof'ev and I. S. Tupitsyn},
  title = {Interaction-enhanced nesting in Spin-Fermion and Fermi-Hubbard models},
  journal = {Phys. Rev. Research},
  volume = {6},
  pages = {L032058},
  year = {2024},
doi = {10.1103/PhysRevResearch.6.L032058}
}

@article{Hidden,
  author = { Shiro Sakai and  Marcello Civelli and Masatoshi Imada},
  title = {Hidden Fermionic Excitation Boosting High-Temperature Superconductivity in Cuprates},
  journal = {Phys. Rev. Lett},
  volume = {116},
  pages = {057003},
  year = {2016},
doi = {10.1103/PhysRevLett.116.057003}
}

@article{Imada,
  author = { M. Imada and T. J. Suzuki},
  title = {Excitons and Dark Fermions as Origins of Mott Gap, Pseudogap and Superconductivity in Cuprate Superconductors - General Concept and Basic Formalism Based on Gap Physics},
  journal = {J. Phys. Soc. Jpn.},
  volume = {88},
  pages = {024701},
  year = {2019},
doi = {10.7566/JPSJ.88.024701}
}

@article{Previous,
  author = {Coleman, Piers and Panigrahi, Aaditya and Tsvelik, Alexei},
  title = {A microscopic model of a fractionalized Fermi liquid},
  journal = {arXiv},
  volume = {2511.01115},
  pages = {},
  year = {},
doi = {10.48550/arXiv.2511.01115}
}

@article{Peter1,
  author = {H.-B. Yang and J. D. Rameau and P. D. Johnson and T. Valla and A. Tsvelik and G. D. Gu},
  title = {Emergence of preformed Cooper pairs from the doped Mott insulating state in Bi$_2$Sr$_2$CaCu$_2$O$_{8+\delta}$},
  journal = {Nature},
  volume = {456},
  pages = {77},
  year = {2008},
doi = {10.1038/nature07400}
}

@misc{AGD,
  title={Methods of quantum field theory in statistical physics},
  author={Abrikosov, Aleksei, Alekseevich and Gorkov, Lev Petrovich and Dzyaloshinski, Igor Ekhielevich and Silverman, Richard A and Weiss, George H},
  year={1964},
  publisher={American Institute of Physics}
}

@article{Fabrizio2023,
  title = {Spin-Liquid Insulators Can Be Landau's Fermi Liquids},
  author = {Fabrizio, Michele},
  journal = {Phys. Rev. Lett.},
  volume = {130},
  issue = {15},
  pages = {156702},
  numpages = {6},
  year = {2023},
  month = {Apr},
  publisher = {American Physical Society},
  doi = {10.1103/PhysRevLett.130.156702},
  url = {https://link.aps.org/doi/10.1103/PhysRevLett.130.156702}
}

@article{Dzyaloshinskii2003,
  title = {Some consequences of the Luttinger theorem: The Luttinger surfaces in non-Fermi liquids and Mott insulators},
  author = {Dzyaloshinskii, Igor},
  journal = {Phys. Rev. B},
  volume = {68},
  issue = {8},
  pages = {085113},
  numpages = {6},
  year = {2003},
  month = {Aug},
  publisher = {American Physical Society},
  doi = {10.1103/PhysRevB.68.085113},
  url = {https://link.aps.org/doi/10.1103/PhysRevB.68.085113}
}

@article{Kuklov,
  title = {Dual view of the Z2-Gauged XY Model in 3D},
  author = {Coleman, Piers and Kuklov, Anatoly and Tsvelik, Alexei M.},
  journal = {Phys. Rev. Lett.},
  volume = {134},
  pages = {236001},
  year = {2025},
  publisher = {American Physical Society},
  doi = {10.1103/PhysRevLett.134.236001},
  url = {https://link.aps.org/doi/10.1103/PhysRevLett.134.236001}
}
\end{document}